\documentclass[journal]{IEEEtran}
\usepackage{amsmath,amsfonts}
\usepackage{amssymb}
\usepackage{algorithm}
\usepackage{array}
\usepackage{textcomp}
\usepackage{url}
\usepackage{verbatim}
\usepackage{xcolor}
\usepackage{graphicx}
\usepackage{setspace}
\usepackage{adjustbox}
\usepackage{multirow}
\usepackage{diagbox}
\usepackage{color}
\usepackage{pifont}
\usepackage{afterpage}
\usepackage{float}
\usepackage{booktabs}
\usepackage{algpseudocode}
\usepackage[caption=false,font=footnotesize,labelfont=rm,textfont=rm]{subfig}

\makeatletter

\newcommand{\Rmnum}[1]{\expandafter\@slowromancap\romannumeral #1@}
\makeatother

\usepackage{cite}
\usepackage[nottoc]{tocbibind}
\usepackage{hyperref}
\begin{document}

\title{Sensing the Variation of Wireless Propagation Environment Based on Channel Knowledge Map}

\author{Haohan Li, Wence Zhang, Xiaoyan Shao, Xu Bao, Jing Xia, Yong Zeng, \textit{Fellow, IEEE}
\thanks{
This work was supported in part by the National Natural Science Foundation of China (Nos. 62571116, 62671280), Natural Science Foundation for Distinguished Young Scholars of Jiangsu Province (BK20240070). 
		
H. Li, W. Zhang, X. Shao, X. Bao and J. Xia are with the School of Computer Science and Communication Engineering, Jiangsu University, Zhenjiang 212000, China, and the Jiangsu Key Laboratory of Security Technology for Industrial Cyberspace, Zhenjiang 212000, China. (e-mails: \{2222308001, 2212308002\}@stmail.ujs.edu.cn; \{wencezhang, xbao, jingxia\}@ujs.edu.cn)
		
Y. Zeng is with the National Mobile Communications Research Laboratory, Southeast University, Nanjing 210096, China, also with the Purple Mountain Laboratories, Nanjing 211111, China (e-mail: yong\_zeng@seu.edu.cn).

\textit{Corresponding author: Wence Zhang.}
}
}

%



\maketitle

\begin{abstract}
With the evolution of sixth-generation (6G) wireless systems, integrated sensing and communication (ISAC) has raised the demand for high-precision environmental perception. To address the limited capability of existing techniques in sensing environmental variations, this work proposes a novel channel knowledge map (CKM)-based sensing framework that enables structural change sensing in the wireless propagation environment. By leveraging historical CKM data and real-time user channel, a mathematical model is formulated to infer the changing reflectors which represents the environment variations. Specifically, we propose the \underline{I}nformation \underline{A}ggregation Based Robust \underline{T}wo-\underline{S}tage \underline{R}eflection \underline{P}oint \underline{R}efinement (IA-TS-RPR) scheme for reflection point (RP) estimation, which first uses block coordinate descent (BCD) and Riemannian manifold optimization (RMO) for initial estimation, then applies virtual base station (VBS) clustering for information aggregation in the sensitive area (SA) to  refine the estimation. 
{\color{black}Furthermore, we propose an \underline{E}nvironment \underline{V}ariation \underline{S}ensing scheme based on the \underline{H}ausdorff-Nearest Neighbor \underline{C}riterion and \underline{T}ime-\underline{W}indow \underline{V}oting (EVS-HC-TWV). Specifically, the Hausdorff distance is used to match each user to its most relevant CKM grid point (GP), and a time-window-based multi-user voting mechanism is further employed to improve robustness against positional deviations.  
Simulation results demonstrate that the proposed IA-TS-RPR scheme maintains sub-meter RP  position estimation accuracy under user positioning errors, significantly outperforming the scheme based on time of arrival (ToA), linear minimum mean square error (LMMSE) and least squares (LS).}
The proposed scheme successfully senses and localizes dynamic reflectors, showing strong generalizability and robustness. 
\end{abstract}


\section{Introduction}

{\color{black}


\IEEEPARstart{T}{he} sixth-generation (6G) wireless systems are expected to exploit higher frequency bands including millimeter wave (mmWave) and terahertz spectrum and denser deployments of antenna arrays to enable the deep integration of wireless sensing and communication functionalities within a single system~\cite{ref_6g}. Integrated sensing and communication (ISAC), which facilitates joint design based on shared hardware or information, has attracted significant attention from both academia and industry \cite{ref10_ISAC}. It enables simultaneous communication and sensing using the same waveform over the same spectral resources, thus to enhance spectral efficiency~\cite{ref2_ISAC} and offer services such as target localization and imaging, which are promising for applications including low-altitude economy, intelligent transportation and smart manufacturing~\cite{ref3_ISAC}.

One of the key components for ISAC is high-precision wireless environment perception~\cite{ref4_ISAC}. On the one hand, wireless environment sensing provides the perceptual foundation at the communication end to support simultaneous localization and mapping (SLAM)~\cite{ref5_ISAC}. On the other hand, the obtained sensing information can be utilized to enhance communication performance, enabling advanced functionalities such as perception-assisted wireless resource pre-allocation \cite{ref6_ISAC} and the construction of wireless channel digital twin models~\cite{ref8_ISAC}.

In recent years, the emerging concept of channel knowledge map (CKM) has established a critical link between the wireless environment and channel information \cite{ref4_CKM}, becoming one of the key enablers for ISAC systems~\cite{ref3_CKM,ref2_CKM}. {\color{black} CKM is a site-specific database or AI model that stores the positions of transmitters and receivers along with the corresponding channel characteristics, which facilitates enhanced environmental awareness and simplifies real-time channel estimation~\cite{ref5_CKM}.} Typical channel knowledge stored in a CKM includes path loss, angle of arrival (AoA),  angle of departure (AoD) and etc. By combining these channel parameters with the location information of transceivers, CKM can be leveraged to infer properties of the wireless propagation environment, such as obstacles and scatterers~\cite{rel9_CKMY}. CKM has been used in a wide range of applications, including pilot overhead reduction~\cite{rel1_CKMY}, position-predictive wireless resource allocation~\cite{rel6_CKMY}, CKM-assisted localization and sensing~\cite{Wu2026TWC,rel7_CKMY,ref_why} and etc. These studies demonstrate that CKM can provide strong environment-aware priors and thereby improve efficiency and robustness in communication and sensing tasks.

{\color{black}
Building on this capability, this work investigates CKM-assisted environmental variation sensing by comparing historical channel knowledge with real-time channel observations from users. Specifically, single-reflection multipath components are leveraged to estimate the locations of reflection points (RPs), such that dynamic variations in surrounding reflectors can be inferred from the discrepancy between the CKM and current channel observations. The main contributions of this work are summarized as follows.

\begin{itemize}
    \item Given real-time channel observations from multiple users and historical channel information stored in the CKM, the problem of CKM-based environmental variation sensing is formulated, which characterizes environmental changes by identifying the set of varying reflectors, thereby enabling the sensing of dynamic variation in the propagation environment.    
    \item  An \underline{I}nformation \underline{A}ggregation Based Robust \underline{T}wo-\underline{S}tage \underline{R}eflection \underline{P}oint \underline{R}efinement (IA-TS-RPR) algorithm is proposed based on the concept of sensitive area (SA). Each reflector is associated with an SA, in which the users receive signals reflected by them. In the first stage, an initial estimation algorithm is designed using block coordinate descent (BCD) and Riemannian manifold optimization (RMO). In the second stage, hierarchical clustering of VBSs is employed to refine the estimated RP positions. The proposed IA-TS-RPR scheme achieves sub-meter RP estimation accuracy under user positioning errors, outperforming the ToA based scheme in \cite{Xie2024ToA}, linear minimum mean square error (LMMSE) and least squares (LS).

    \item The scheme of \underline{E}nvironment \underline{V}ariation \underline{S}ensing based on \underline{H}ausdorff-Nearest Neighbor (HNN) \underline{C}riterion and \underline{T}ime-\underline{W}indow \underline{V}oting (EVS-HC-TWV) is proposed, in which the Hausdorff distance is employed to quantify the discrepancy between real-time channel observations from multiple users and historical channel information stored in the CKM, enabling the association between users and the most relevant CKM grid points (GPs). To mitigate the randomness in judgments made by individual users, a collaborative voting scheme based on a time-window is further employed to enhance the accuracy and robustness of environmental variation sensing.
    
    \item Comprehensive simulations validate the effectiveness of the proposed IA-TS-RPR and EVS-HC-TWV schemes. The proposed IA-TS-RPR achieves sub-meter accuracy for over 90\% of the RP estimates. The proposed EVS-HC-TWV achieves over 95\% sensing accuracy across diverse scenarios, demonstrating remarkable robustness and generalizability.
\end{itemize}

The remainder of this paper is organized as follows. Section~II reviews related works on target sensing and physical environment sensing. Section~III presents the system model and formulates the problem of environmental variation sensing based on CKM. Section~IV introduces the proposed two-stage robust RP position algorithm, i.e., IA-TS-RPR, and Section~V describes the scheme of environment variation sensing based on \underline{H}ausdorff-\underline{N}earest \underline{N}eighbor (HNN) criterion and time-window voting, i.e., EVS-HC-TWV. Numerical results and discussion are provided in Section~VI. Conclusions are drawn in Section~VII.

{\textit{Notations:} ${\mathbb{C}}^{{M}\times {N}}$ represents the set of $M \times N$ complex-valued matrices; $\otimes$ denotes the Kronecker product; $\mathbf{X}^\text{T}$, $\mathbf{X}^\text{-1}$, $\mathbf{X}^{\dagger}$ denote the transpose, inverse and pseudo-inverse of matrix $\mathbf{X}$, respectively; ${\left\| \cdot \right\|}_2$ represents the Euclidean norm; $\left| \cdot \right|$ is the absolute value of a complex number; $\mathcal{C}\mathcal{N}(\mu,\sigma^{2})$ denotes a circularly symmetric complex Gaussian distribution with mean $\mu$ and variance $\sigma^{2}$;  $\bigcup$ and $\bigcap$ represent the union and intersection of sets, respectively.}}

\section{Related Works}
Existing studies on environment sensing can typically be categorized by the sensing objective: (i) targets in the environment \cite{xiao2024TSP,Keskin2024TWC,hu2024TWC,Pegoraro2024TWC,Xie2024ToA,venus2024graph,yan2025TGCN,Li2025TVT}, and (ii) the physical environment itself \cite{rel2_SLAM,rel4_SLAM,rel5_SLAM,rel3_SLAM,rel8_SLAM}.

Conventional techniques for target detection in ISAC systems are usually based on wireless echo signals in mono-static systems \cite{xiao2024TSP,Keskin2024TWC}. Using a multiple-input multiple-output (MIMO) orthogonal frequency-division multiplexing (OFDM) technique, the work in \cite{xiao2024TSP} jointly estimated angle, range and velocity from echoes within a coherent processing interval. To improve the target detection performance in high-mobility scenarios,  a generalized likelihood ratio test (GLRT)-based multi-target detection and delay–Doppler–angle estimation scheme are proposed in  \cite{Keskin2024TWC}, which improved detection capability of the target range and velocity. In bi-static ISAC systems, target detection is generally based on the cooperation between the transmitter and the receiver \cite{hu2024TWC, Pegoraro2024TWC}. The work in  \cite{hu2024TWC} derives performance bounds from channel state information (CSI) ratios and optimizes waveforms, thereby improving Doppler sensing accuracy. Considering practical challenges, another work in \cite{Pegoraro2024TWC} leverages inter-packet channel correlation to compensate for timing offsets and track multipath reflections using a dynamically selected reference path, which is able to eliminate frequency offsets and improve tracking performance and micro-Doppler quality. { \it However, these methods aim to detect the beam angle or the range of the potential targets rather than the positions, which are usually difficult to determine without the help of LoS paths in these schemes.}

To take non-line-of-sight (NLoS) channels into consideration, prior works often exploited multipath component (MPC) features for target localization, including time of arrival (ToA) \cite{Xie2024ToA,venus2024graph} and angular information such as AoA/AoD \cite{yan2025TGCN,Li2025TVT}. For ToA-based sensing, an indoor NLoS based scatterer localization method was proposed in \cite{Xie2024ToA}, which converts the localization with NLoS paths into an equivalent line-of-sight (LoS) problem and achieved centimeter-level accuracy. Another work in \cite{venus2024graph} proposes a factor graph framework with a particle-based sum-product algorithm for target localization by leveraging a statistical multipath model, which  remains accurate even under fully obstructed line-of-sight (OLoS) conditions.  {\it However, these ToA based sensing schemes relied on accurate clock synchronization.} To tackle this challenge, AoD are further used with ToA in \cite{yan2025TGCN} and \cite{Li2025TVT} for jointly target position and clock offset estimation, where a reconfigurable intelligent surface (RIS) was introduced which helps to improved the estimation performance through phase optimization. {\it However, this method involves additional hardware and the  high computational complexity degrades its real-time performance.}

Regarding physical environment sensing, a typical application is related to wireless SLAM. Existing works usually represent the environment using virtual points \cite{rel2_SLAM,rel4_SLAM,rel5_SLAM,rel3_SLAM,rel8_SLAM}. A virtual point is the image of the base station (BS) (or the access point) with respect to a reflecting surface, usually termed as virtual anchors (VAs) or virtual BSs (VBSs). Early work in \cite{rel2_SLAM} introduces a Channel-SLAM approach based on Bayesian filtering, where multipath components are modeled as static VBS to fuse multipath delay information, enabling joint localization of a mobile receiver and estimation of RPs in static environments. Similarly, a belief propagation based SLAM algorithm  is proposed in \cite{rel4_SLAM} to jointly estimate the agent trajectory and the positions of VBSs. To resolve the ambiguity of associating multiple observations with a single VBS due to practical imperfections \cite{rel2_SLAM,rel4_SLAM}, the work in \cite{rel5_SLAM} proposes a particle belief propagation strategy for joint association of multiple measurements, which is further fused to improve the VBS localization performance. Building on the above single-bounce propagation paths \cite{rel2_SLAM,rel4_SLAM,rel5_SLAM}, the master-VBS strategy in \cite{rel3_SLAM} further accounts for double-bounce propagation by adopting reflecting-surface modeling with ray tracing to evaluate propagation-path availability, thereby improving positioning accuracy in wireless SLAM. Furthermore, to accommodate multi-user cooperation and broaden applicability, a plug-and-play and crowdsourced SLAM framework was developed to handle measurement errors and bias, achieving high-precision localization without prior knowledge of BS positions~\cite{rel8_SLAM}. {\color{black}In addition, the work in \cite{Baquero2022TVT} investigates mmWave mobile sensing and environment mapping with ray-tracing and measurement validation.} {\it However, existing works focus on static environment sensing and the computational complexity is high due to the employment of complex statistical inference methods.  Sensing the variation of the environment is still an open problem. }

In summary, prior studies demonstrate that multipath information can support target and environment sensing. Existing sensing methods based on ToA typically require synchronization mechanisms or joint estimation of clock offsets, which increases the system complexity, while those environment sensing schemes do not consider the variation of the environment itself and usually exhibit high complexity. In fact, CKM can provide extensive environment-related multipath information, and thus can support environment sensing tasks. Moreover, it works as a database and preprocessing on the environment information can be done offline, which helps to reduce the real-time application complexity.  By comparing the CKM data with the instantaneous channel information on the user side,  it is feasible to estimate the change of the wireless propagation environment.

\section{System Model and Problem Formulation}
{\color{black}This section first introduces the system model, including the Saleh–Valenzuela (S-V) channel model and the structure of the CKM. Then, a mathematical formulation of the CKM-based environmental perception problem is developed.} 

\subsection{System Model}
{\color{black}
Consider a single-cell mmWave communication system, as illustrated in Fig.~\ref{fig3.1}. The BS is located at ${{\mathbf{p}}^{\text{b}}}={{\left[ {{x}^{\text{b}}},{{y}^{\text{b}}},{{z}^{\text{b}}} \right]}^{\text{T}}}$ and is equipped with ${{N}_{\text{B}}}$ antennas. The BS serves $K$ users, each equipped with ${{N}_{\text{U}}}$ antennas. The position of user $k$ is denoted by ${\mathbf{p}}_{k}^{\text{u}}={{\left[ x_{k}^{\text{u}},y_{k}^{\text{u}},z_{k}^{\text{u}} \right]}^{\text{T}}}$, which can be obtained using ISAC technology. Therefore, ${{\mathbf p}^{\text b}}$  and ${{\mathbf p}^{\text u}_k}$ are both assumed to be available at the BS \cite{ref4_ISAC}.}
There exist ${{M}_{0}}$ reflectors in the environment. The position of reflector $m$ ($m = 1, \ldots, M_0$) is denoted by $\mathbf{p}_{m}^{\text{S}}={{\left[ x_{m}^{\text{S}},y_{m}^{\text{S}},z_{m}^{\text{S}} \right]}^{\text{T}}}$, and its physical attributes (such as size and material type)  which affect wireless signal propagation are represented by the set ${{\Omega}_{m}}$. Let ${{S}_{m}} = \left\{ \mathbf{p}_{m}^{\text{S}}, {{\Omega}_{m}} \right\}$ denote the information set of reflector $m$. {\color{black}In this work, we only focus on sensing the variation of reflector positions ${\mathbf{p}_{m}^{\text{S}}}$,  while estimating ${\Omega}_{m}$ is left for future work.} Therefore, the complete set of reflector information in the environment is defined as ${E}^0 = \left\{ {{S}_{m}} \mid m = 1, \ldots, M_0 \right\}$. These reflectors significantly affect the channel characteristics between the BS and the users.

\begin{figure}[!thb]
	\centering
	{\includegraphics[width=.7\linewidth]{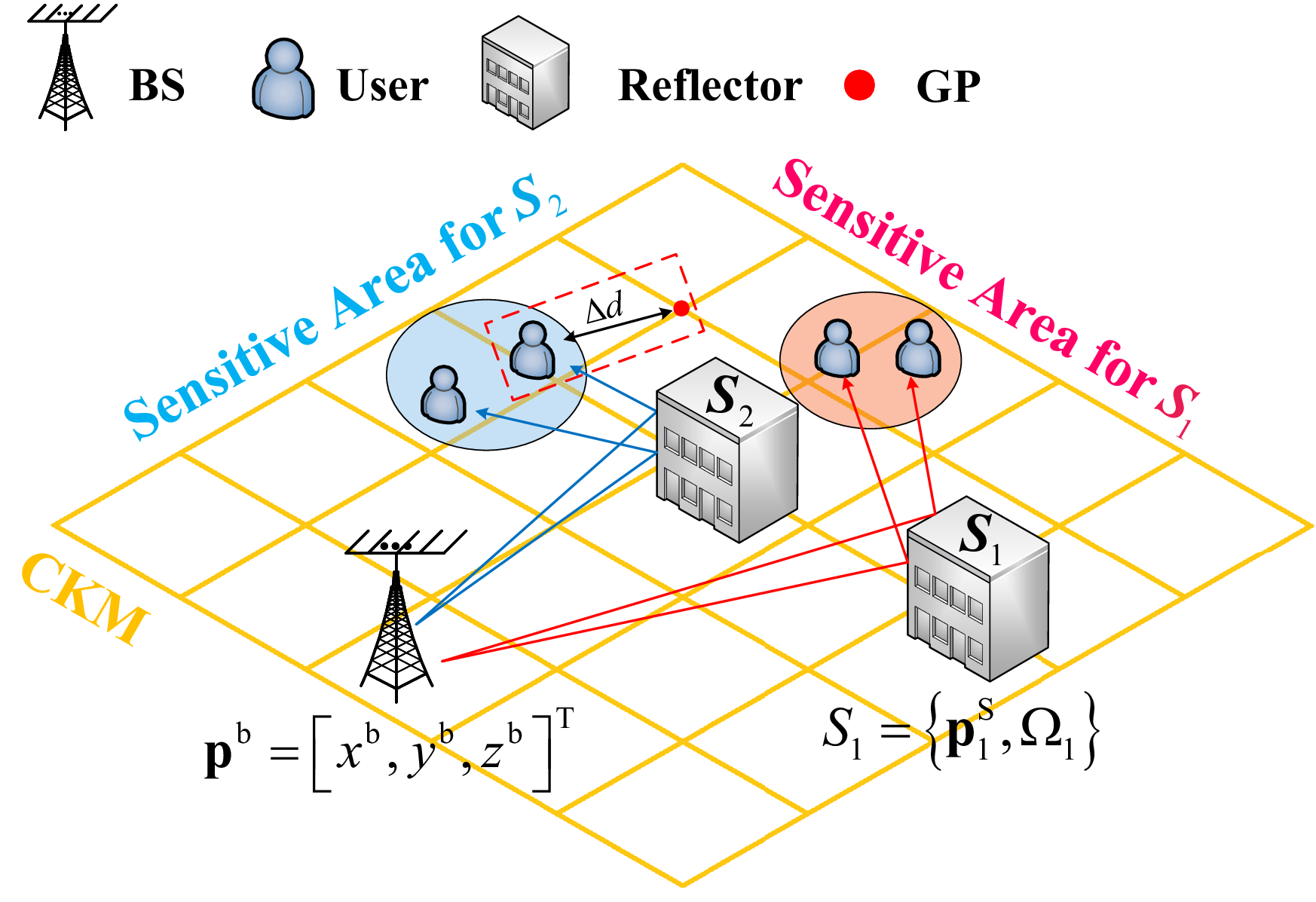}}
	\caption{An illustration of  CKM based environmental variation sensing. There are two reflectors labeled as ${S}_{1}$ and ${S}_{2}$, respectively. Each of them corresponds to a SA, in which the users receive signals reflected by them.  A CKM is constructed with $N_{\text{CKM}}$ GPs (i.e., the yellow grids).  The user positions may deviate from the reference GP (i.e., the red point). }
	\label{fig3.1}
\end{figure}

In this work, we adopt an extended S-V geometric channel model to characterize mmWave channels \cite{ref_SV}. Taking the uplink as an example, the channel matrix from user $k$ to the BS is denoted by ${{\mathbf{H}}_{k}} \in {{\mathbb{C}}^{{{N}_{\text{B}}}\times {{N}_{\text{U}}}}}$ and is expressed as
\begin{equation}\label{equ1.1}
{{\mathbf{H}}_{k}}=\sqrt{\frac{{{N}_{\text{U}}}{{N}_{\text{B}}}}{{{L}_{k}}}}\sum\limits_{i=1}^{{{L}_{k}}}{{{\alpha }_{k,i}}{{\mathbf{a}}_{\text{r}}}\left( \theta _{k,i}^{\text{A}},\phi _{k,i}^{\text{A}} \right)\mathbf{a}_{\text{t}}^{\text{H}}\left( \theta _{k,i}^{\text{D}},\phi _{k,i}^{\text{D}} \right)},
\end{equation}
where $L_k$ is the number of propagation paths; $\alpha_{k,i}$ denotes the gain of the $i$-th path and follows a circularly symmetric complex Gaussian distribution with zero mean and variance $\sigma _{k,i}^{2}$, i.e., ${{\alpha }_{k,i}}\sim \mathcal{C}\mathcal{N}(0,\sigma _{k,i}^{2})$. $\theta _{k,i}^{\text{A}}$ and $\phi _{k,i}^{\text{A}}$ represent the elevation and azimuth AoA, while $\theta _{k,i}^{\text{D}}$ and $\phi _{k,i}^{\text{D}}$ denote the elevation and azimuth AoD, where $\theta _{k,i}^{\text{A}},\theta _{k,i}^{\text{D}} \in \left[ -\frac{\pi}{2}, \frac{\pi}{2} \right]$, and $\phi _{k,i}^{\text{A}},\phi _{k,i}^{\text{D}} \in \left[ -\pi, \pi \right]$. ${{\mathbf{a}}_{\text{r}}}(\theta _{k,i}^{\text{A}},\phi _{k,i}^{\text{A}})$ and ${{\mathbf{a}}_{\text{t}}}(\theta _{k,i}^{\text{D}},\phi _{k,i}^{\text{D}})$ denote the array response vectors at the receiver and transmitter, respectively.
Assuming that a uniform planar array (UPA) is used where the antenna elements are arranged uniformly in a matrix on the X–Y plane with a distance of $d$, and the numbers of elements in the horizontal and vertical directions are $N_{\text{H}}$ and $N_{\text{V}}$, respectively, the corresponding array response vectors are given by
\begin{equation}\label{equ1.2}
{{\mathbf{a}}_{\text{H}}}=\frac{1}{\sqrt{{{N}_{\text{H}}}}}{{\left[ 1,{{\text{e}}^{-\text{j}\frac{2\pi }{\lambda }d\cos \theta \sin \phi }},\cdots ,{{\text{e}}^{-\text{j}\left( {{N}_{\text{H}}}-1 \right)\frac{2\pi }{\lambda }d\cos \theta \sin \phi }} \right]}^{\text{T}}},
\end{equation}
\begin{equation}\label{equ1.3}
{{\mathbf{a}}_{\text{V}}}=\frac{1}{\sqrt{{{N}_{\text{V}}}}}{{\left[ 1,{{\text{e}}^{-\text{j}\frac{2\pi }{\lambda }d\sin \phi }},\cdots ,{{\text{e}}^{-\text{j}\left( {{N}_{\text{V}}}-1 \right)\frac{2\pi }{\lambda }d\sin \phi }} \right]}^{\text{T}}},
\end{equation}
where $\lambda$ denotes the wavelength. 
{\color{black}Therefore, the array response vector of the UPA is ${\mathbf a}_{\text{UPA}} (\phi,\theta) = {\mathbf{a}}_{\text{V}}\otimes {{\mathbf{a}}_{\text{H}}}$, i.e., 
\begin{equation}\label{equ1.4}
\begin{aligned}
   &{{\mathbf{a}}_{\text{UPA}}}\left( \phi ,\theta  \right) =\\
  & \frac{1}{\sqrt{N_{\mathrm{H}}N_{\mathrm{V}}}}
\Big[
1,\ \cdots,\ 
{\text{e}}^{-\text{j}\frac{2\pi d}{\lambda}\big((N_{\mathrm{H}}-1)\cos\theta\sin\phi + (N_{\mathrm{V}}-1)\sin\phi\big)}
\Big]^{\mathrm{T}} .
\end{aligned}
\end{equation}}

As shown in Fig.~\ref{fig3.1} and (\ref{equ1.1}), for the ${{l}_{k}}$-th path ($1\le {{l}_{k}}\le {{L}_{k}}$) of user $k$, the key channel parameters including AoD/AoA and the path gain are represented by  
$\mathbf{r}_{k,{{l}_{k}}}^{\text{u}}={{[\phi _{k,{{l}_{k}}}^{\text{D}},\theta _{k,{{l}_{k}}}^{\text{D}},\phi _{k,{{l}_{k}}}^{\text{A}},\theta _{k,{{l}_{k}}}^{\text{A}},{{\alpha }_{k,{{l}_{k}}}}]}^{\text{T}}}$.  
Accordingly, the complete channel information for user $k$ can be expressed as  
$\mathbf{R}_{k}^{\text{u}}=\{\mathbf{r}_{k,1}^{\text{u}},\cdots ,\mathbf{r}_{k,{{l}_{k}}}^{\text{u}},\cdots ,\mathbf{r}_{k,{{L}_{k}}}^{\text{u}}\}$. $\mathbf{r}_{k,{{l}_{k}}}^{\text{u}}$ is affected by environmental variations and user mobility. However, in quasi-static scenarios where the user remains stationary or moves slowly and the surrounding environment changes gradually, $\mathbf{r}_{k,{{l}_{k}}}^{\text{u}}$ mainly depends on the user’s position $\mathbf{p}_{k}^{\text{u}}$. Under this condition, given the environmental reflector information ${E}^{0}$, $\mathbf{r}_{k,{{l}_{k}}}^{\text{u}}$ can be expressed as a function of $\mathbf{p}_{k}^{\text{u}}$, that is,  
\begin{equation}\label{equ1.5}
	\mathbf{r}_{k,{{l}_{k}}}^{\text{u}}=\mathcal{W}\left( \mathbf{p}_{k}^{\text{u}}|{{{E}^0}} \right),
\end{equation}
where $\mathcal{W}\left( \cdot  \right)$ denotes the latent mapping from spatial coordinates to channel information, which depends on factors such as the operating frequency and propagation environment.  However,  the function $\mathcal{W}\left( \cdot \right)$ in (\ref{equ1.5}) is generally intractable to represent in closed form. {\color{black}Therefore, it is advocated to  be approximated using CKM \cite{ref5_CKM},  which can be constructed in advance using methods in \cite{Shao2026IOT,rel3_CKM} and is assumed  available in this work.}

As shown in Fig.~1, the coverage area is uniformly divided into ${{N}_{\text{CKM}}}$ GPs. The position and corresponding channel knowledge of each GP are recorded. In this case, the CKM can be represented by the set ${{\mathcal{R}}_{\text{C}}}$, given by  {\color{black}
\begin{equation}\label{equ1.6}
	{{\mathcal{R}}_{\text{C}}}=\left\{ \left\langle \mathbf{p}_{1}^{\text{c}},\mathbf{R}_{1}^{\text{c}} \right\rangle ;\left\langle \mathbf{p}_{2}^{\text{c}},\mathbf{R}_{2}^{\text{c}} \right\rangle ;\ldots ;\left\langle \mathbf{p}_{{{N}_{\text{CKM}}}}^{\text{c}},\mathbf{R}_{{{N}_{\text{CKM}}}}^{\text{c}} \right\rangle  \right\}.
\end{equation}

For example, consider the $n$-th GP ($1 \le n \le {N_{{\rm{CKM}}}}$) located at $\mathbf{p}_{n}^{\text{c}} = {{\left[ x_{n}^{\text{c}}, y_{n}^{\text{c}}, z_{n}^{\text{c}} \right]}^{\text{T}}}$. Its corresponding channel knowledge $\mathbf{R}_{n}^{\text{c}}$ can be expressed as  
\begin{equation}\label{equ1.7}
	\mathbf{R}_{n}^{\text{c}}=\left[ \begin{matrix}
		\phi _{n,1}^{\text{D}} & \theta _{n,1}^{\text{D}} & \phi _{n,1}^{\text{A}} & \theta _{n,1}^{\text{A}} & {{\alpha }_{n,1}}  \\
		\vdots  & \vdots  & \vdots  & \vdots  & \vdots   \\
		\phi _{n,{{l}_{n}}}^{\text{D}} & \theta _{n,{{l}_{n}}}^{\text{D}} & \phi _{n,{{l}_{n}}}^{\text{A}} & \theta _{n,{{l}_{n}}}^{\text{A}} & {{\alpha }_{n,{{l}_{n}}}}  \\
		\vdots  & \vdots  & \vdots  & \vdots  & \vdots   \\
		\phi _{n,{{L}_{n}}}^{\text{D}} & \theta _{n,{{L}_{n}}}^{\text{D}} & \phi _{n,{{L}_{n}}}^{\text{A}} & \theta _{n,{{L}_{n}}}^{\text{A}} & {{\alpha }_{n,{{L}_{n}}}}  \\
	\end{matrix} \right],
\end{equation} 
where $L_n$ denotes the number of valid propagation paths at this GP. For the ${l_n}$-th path ($1 \le l_n \le L_n$), $\phi _{n,{{l}_{n}}}^{\text{D}}$ and $\theta _{n,{{l}_{n}}}^{\text{D}}$ represent the azimuth and elevation AoD, respectively; $\phi _{n,{{l}_{n}}}^{\text{A}}$ and $\theta _{n,{{l}_{n}}}^{\text{A}}$ denote the azimuth and elevation AoA, respectively; and ${{\alpha }_{n,{{l}_{n}}}}$ represents the path gain. Note that the perfect value of $\alpha_{n,i}$ is difficult to obtain due to hardware and other practical issues. Instead, it is reasonable to store its statistics, such as the mean value and/or  the variance. Here it is only used to maintain consistency between the system model and the CKM description. They are not necessary in the algorithms proposed in the following sections.}

\subsection{Problem Formulation}
Due to the high sensitivity of mmWave signals to path blockage, the emergence or vanishing of reflectors in the environment can directly alter channel characteristics such as the AoA and AoD of propagation paths. Therefore, subtle changes in angular information (e.g., AoA and AoD) can be exploited to sense variations in environmental reflectors. Let the set of reflector information after environmental changes be denoted as {${E}^1 = \left\{ S_{m} \mid m = 1, \ldots, M_{1} \right\}$}. Then, the set of changed reflectors can be expressed as the symmetric difference between {${E}^1$ and ${E}^0$:
$\Delta {E}\triangleq {{{E}^0}}\Delta {{{E}^1}}=\left( {{{E}^0}}\bigcup {{{E}^1}} \right)-\left( {{{E}^0}}\bigcap {{{E}^1}} \right)$}. This set includes all reflectors that have changed, and it represents the structural variation in the environment relative to the initial state.

{\color{black}Given the current set of user channel observations ${{\mathcal{R}}_{\text{U}}}=\left\{ \left\langle \mathbf{p}_{1}^{\text{u}},\mathbf{R}_{1}^{\text{u}} \right\rangle ;\left\langle \mathbf{p}_{2}^{\text{u}},\mathbf{R}_{2}^{\text{u}} \right\rangle ;\cdots \text{ }\cdots ;\left\langle \mathbf{p}_{K}^{\text{u}},\mathbf{R}_{K}^{\text{u}} \right\rangle  \right\}$
and the CKM ${{\mathcal{R}}_{\text{C}}}$, this work aims to infer the environmental reflector variation $\Delta {E}$.} The problem can be formulated as
\begin{equation}\label{equ1.8}
	\left( \text{P1} \right)\quad \widehat{\Delta {E}}=\arg \underset{\Delta {E}}{\mathop{\max }}\,\quad P\left( \Delta {E}\left| {{\mathcal{R}}_{\text{C}}}, \right.{{\mathcal{R}}_{\text{U}}},{{E}^0} \right),
\end{equation}
where $\widehat{\Delta {E}}$ denotes the estimated set of varying reflectors, and the expression represents the probability that a given $\Delta {E}$ has occurred, conditioned on $\mathcal{R}_{\text{C}}$, $\mathcal{R}_{\text{U}}$, and ${E}_{0}$. This problem can thus be interpreted as identifying the most likely set of environmental reflector changes that explain the observed channel discrepancies, enabling the perception of hidden structural variations in the environment.

There are three major challenges in solving problem \hyperref[equ1.8]{(P1)}.  
\textbf{Challenge 1}: The user-estimated channel information set ${{\mathcal{R}}_{\text{U}}}$ is subject to random errors, and the estimation accuracy significantly affects the reliability of environmental perception. For instance, when the propagation distance is long, even small errors in AoA or AoD can lead to large deviations in the estimated reflector positions.  
\textbf{Challenge 2}: The inference of $\Delta {E}$ relies on the comparison between ${{\mathcal{R}}_{\text{C}}}$ and ${{\mathcal{R}}_{\text{U}}}$; however, the user’s position may not coincide exactly with any GP in ${{\mathcal{R}}_{\text{C}}}$ (as shown in Fig. \ref{fig3.1}), and thus inherent discrepancies exist in their corresponding channel information. These differences may stem either from spatial mismatches or from actual changes in the reflectors, making them difficult to distinguish.  
\textbf{Challenge 3}: The channel information of a single user exhibits randomness, which reduces the accuracy and robustness of detecting reflector variations.

To address the aforementioned challenges, this paper proposes a CKM-based framework for environmental variation sensing. To handle \textbf{Challenge 1}, a two-stage robust RP estimation is developed based on SA information aggregation, which significantly improves the estimation accuracy. For \textbf{Challenge 2}, we devise a user-GP matching strategy based on the HNN criterion to effectively mitigate the risk of misjudgment caused by spatial mismatches. Regarding \textbf{Challenge 3}, a time-window-based multi-user voting mechanism is introduced to enhance the accuracy and robustness of reflector variation sensing.

{\color{black}The operation procedure of the proposed scheme is summarized in Fig.~\ref{fig:ckm_workflow}.} {\color{black}As illustrated in Fig.~\ref{fig:ckm_workflow}, the CKM is constructed in advance using existing methods~\cite{ref4_CKM,ref5_CKM,rel7_CKMY,rel3_CKM}, and its VBS distribution features are extracted as the historical reference using the proposed IA-TS-RPR scheme in Section IV. During online operation, the corresponding user-side VBS features are obtained from real-time channel information, and environmental variation sensing is performed using the proposed EVS-HC-TWV scheme in Section V. Note that the SA is not treated as additional prior information and does not require explicit SA estimation. }
\begin{figure}[!thb]
	\centering
	{\includegraphics[width=\linewidth]{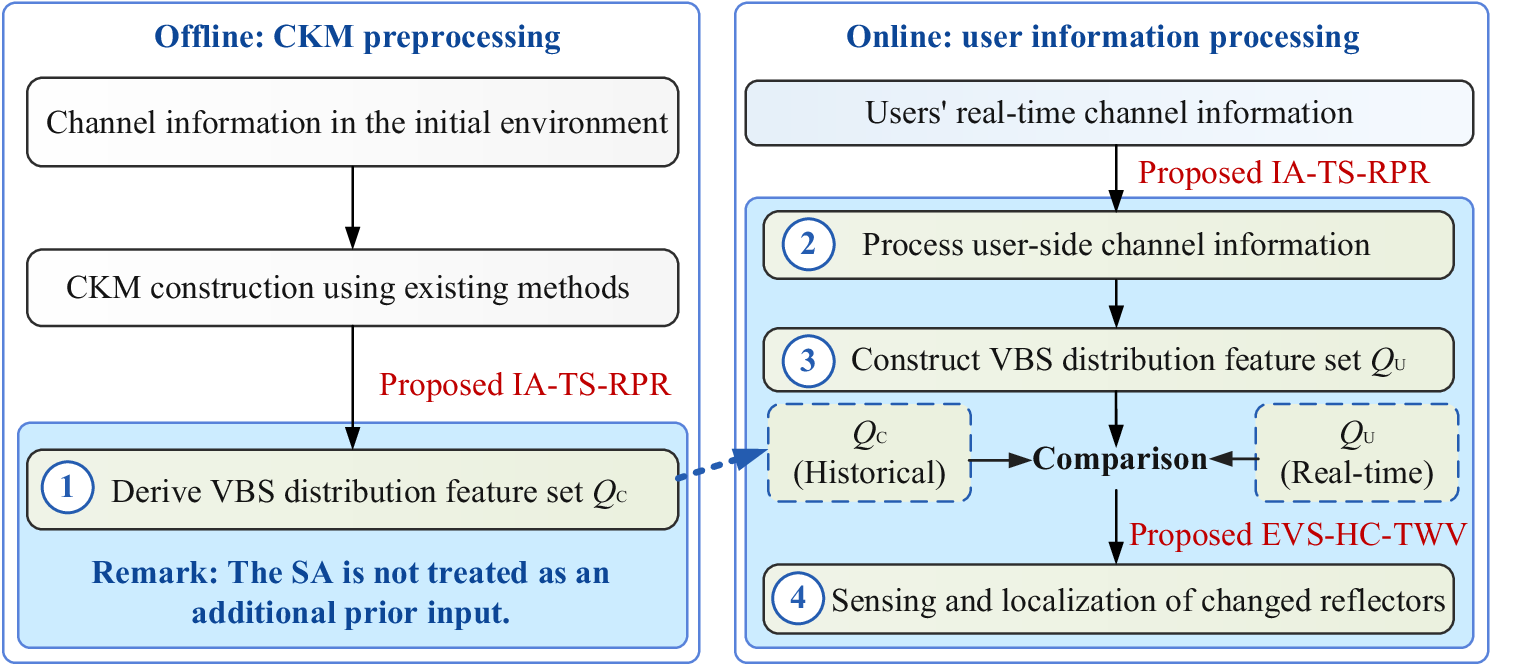}}
	\caption{Initialization and operation procedure of the proposed CKM-based environmental variation sensing framework.}
    \label{fig:ckm_workflow}
\end{figure}

\section{Two-Stage Robust RP Estimation Based on SA Information Aggregation}
Environmental variation sensing can be realized by detecting changes in the RPs along signal propagation paths. Given the AoD and AoA information of a path, two lines can be determined, and the RP is ideally located at their intersection. However, in practice, these angular parameters inevitably contain estimation errors. As illustrated in Fig.~\ref{fig3.2}, taking the elevation angle as an example, the presence of angle estimation errors causes the AoD and AoA lines to no longer intersect precisely, making the RP location ambiguous. Under such conditions, the RP may lie anywhere within the shaded region. Moreover, when the propagation distance is large, even small angular errors can result in significant deviations in the estimated position of the reflector.

\begin{figure}[!thb]
	\centering
	{\includegraphics[width=0.7\linewidth]{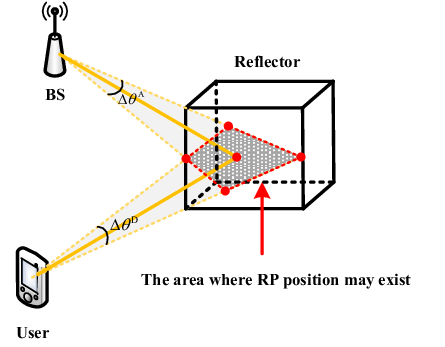}}
	\caption{Illustration of the possible region of the RP under noise. Taking elevation angle $\theta$ as an example, the deviation of the angle between incident and reflected signals leads to non-unique RP positions. The shaded area indicates where the RP may exist.}

	\label{fig3.2}
\end{figure}

This work considers single-reflection NLoS paths \cite{rel3_SLAM}. The method for distinguishing LoS and NLoS components can be found in \cite{rel3_CKM}. Consider a reference path $l$ associated with user $k$, with its angular information given by the set $\{{{\hat{\theta }}^{\text{D}}},{{\hat{\phi }}^{\text{D}}},{{\hat{\theta }}^{\text{A}}},{{\hat{\phi }}^{\text{A}}}\}$, where
\begin{equation}\label{equ1.9}
	\begin{aligned}
		& {{{\hat{\theta }}}^{\text{D}}}={{\theta }^{\text{D}}}+\Delta {{\theta }^{\text{D}}},{{{\hat{\phi }}}^{\text{D}}}={{\phi }^{\text{D}}}+\Delta {{\phi }^{\text{D}}}, \\ 
		& {{{\hat{\theta }}}^{\text{A}}}={{\theta }^{\text{A}}}+\Delta {{\theta }^{\text{A}}},{{{\hat{\phi }}}^{\text{A}}}={{\phi }^{\text{A}}}+\Delta {{\phi }^{\text{A}}}, \\ 
	\end{aligned}
\end{equation}
in which $\hat{\theta}^{\text{D}}$ and $\hat{\phi}^{\text{D}}$ denote the observed elevation and azimuth of AoD, and $\Delta \theta^{\text{D}}$, $\Delta \phi^{\text{D}}$ are the corresponding estimation errors. Similarly, $\hat{\theta}^{\text{A}}$ and $\hat{\phi}^{\text{A}}$ denote the observed elevation and azimuth of AoA, with $\Delta \theta^{\text{A}}$, $\Delta \phi^{\text{A}}$ being the respective estimation errors. These errors are modeled as Gaussian random variables that follow $\mathcal{CN}(0,\sigma^2)$.

To mitigate the estimation errors caused by observation noise, this paper proposes a two-stage RP estimation based on SA information aggregation. In the first stage, each user performs a coarse estimation of the RP position for each path using the path information $\mathbf{r}_{k}^{\text{u}}$. In the second stage, VBSs are introduced, and the coarse RP estimates are used to infer the corresponding VBS location for each path. These VBS locations are then clustered to aggregate information from the same SA, and the clustering results are used to refine the RP estimates for each path. The above two stages are iteratively executed until convergence is achieved.

\subsection{Stage 1: Single-Path RP Estimation Based on BCD-RMO}
Let the true departure and arrival direction vectors of the reference path be denoted by ${{\mathbf{v}}_{\text{D}}}$ and ${{\mathbf{v}}_{\text{A}}}$, respectively, and expressed as follows~\cite{ref_V}  
\begin{equation}\label{equ1.10}
	{{\mathbf{v}}_{\text{D}}}={{\left[ \cos {{\theta }^{\text{D}}}\cos {{\phi }^{\text{D}}},\cos {{\theta }^{\text{D}}}\sin {{\phi }^{\text{D}}},\sin {{\theta }^{\text{D}}} \right]}^{\text{T}}},
\end{equation}
\begin{equation}\label{equ1.11}
	{{\mathbf{v}}_{\text{A}}}={{\left[ \cos {{\theta }^{\text{A}}}\cos {{\phi }^{\text{A}}},\cos {{\theta }^{\text{A}}}\sin {{\phi }^{\text{A}}},\sin {{\theta }^{\text{A}}} \right]}^{\text{T}}}.
\end{equation}

As illustrated in Fig. \ref{fig3.2}, the RP position can be obtained by computing the intersection point of the lines determined by ${{\mathbf{v}}_{\text{D}}}$ and ${{\mathbf{v}}_{\text{A}}}$. However, their observed counterparts ${{\hat{\mathbf{v}}}_{\text{D}}}$ and ${{\hat{\mathbf{v}}}_{\text{A}}}$ are subject to errors, and the corresponding lines may not intersect. Therefore, this paper estimates the RP position by solving for the closest points between the two lines. The resulting optimization problem is formulated as
\begin{equation}\label{equ1.12}
	\left( \text{P}2 \right)\quad \underset{{{q}_{\text{D}}},{{q}_{\text{A}}}\in \mathbb{R}}{\mathop{\min }}\,\quad {{\left\| \left( {{\mathbf{p}}^{\text{u}}}+{{q}_{\text{D}}}{{\mathbf{v}}_{\text{D}}} \right)-\left( {{\mathbf{p}}^{\text{b}}}+{{q}_{\text{A}}}{{\mathbf{v}}_{\text{A}}} \right) \right\|}_2^{2}},
\end{equation}
where $q_{\text{D}}$ denotes the displacement weight along the AoD direction from the user position, and $q_{\text{A}}$ denotes the displacement weight along the AoA direction from the BS position. It is straightforward to verify that problem \hyperref[equ1.12]{(P2)} is a convex optimization problem, and the optimal solution can be obtained by taking the derivative, given by  
\begin{equation}\label{equ1.13}
	\left[ \begin{matrix}
		q_{\text{D}}^{*}  \\
		q_{\text{A}}^{*}  \\
	\end{matrix} \right]={{\left[ \begin{matrix}
				{{\left\| {{\mathbf{v}}_{\text{D}}} \right\|}^{2}} & -\mathbf{v}_{\text{D}}^{\text{T}}{{\mathbf{v}}_{\text{A}}}  \\
				-\mathbf{v}_{\text{D}}^{\text{T}}{{\mathbf{v}}_{\text{A}}} & {{\left\| {{\mathbf{v}}_{\text{A}}} \right\|}^{2}}  \\
			\end{matrix} \right]}^{\dagger }}\left[ \begin{matrix}
		-\mathbf{v}_{\text{D}}^{\text{T}}\left( {{\mathbf{p}}^{\text{u}}}-{{\mathbf{p}}^{\text{b}}} \right)  \\
		\mathbf{v}_{\text{A}}^{\text{T}}\left( {{\mathbf{p}}^{\text{u}}}-{{\mathbf{p}}^{\text{b}}} \right)  \\
	\end{matrix} \right].
\end{equation}

The midpoint of the closest points on the two lines is taken as the estimated RP position, given by  
\begin{equation}\label{equ1.14}
{{\mathbf{p}}^{\text{r}}}=\frac{1}{2}\left[ \left( {{\mathbf{p}}^{\text{u}}}+q_{\text{D}}^{*}{{\mathbf{v}}_{\text{D}}} \right)-\left( {{\mathbf{p}}^{\text{b}}}+q_{\text{A}}^{*}{{\mathbf{v}}_{\text{A}}} \right) \right].
\end{equation}

The estimation for ${{\mathbf{v}}_{\text{D}}}$ and ${{\mathbf{v}}_{\text{A}}}$ is introduced below. Under the framework of maximum likelihood estimation, given the true direction vectors ${{\mathbf{v}}_{\text{D}}}$ and ${{\mathbf{v}}_{\text{A}}}$, the likelihood of observing ${{\hat{\mathbf{v}}}_{\text{D}}}$ and ${{\hat{\mathbf{v}}}_{\text{A}}}$ is given by  
\begin{equation}\label{equ1.15}
	\begin{aligned}
		&P\left( {{{\hat{\mathbf{v}}}_{\text{D}}}},{{{\hat{\mathbf{v}}}_{\text{A}}}}|{{\mathbf{v}}_{\text{D}}},{{\mathbf{v}}_{\text{A}}} \right)  
		\overset{(\text{a})}{\mathop{=}}\,P\left( {{{\hat{\mathbf{v}}}_{\text{D}}}}|{{\mathbf{v}}_{\text{D}}} \right)P\left( {{{\hat{\mathbf{v}}}_{\text{A}}}}|{{\mathbf{v}}_{\text{A}}} \right) \\ 
		&=\frac{1}{{{\left( 2\pi {{\sigma }^{2}} \right)}^{3}}}\exp \left( -\frac{{{\left\| {{{\hat{\mathbf{v}}}}_{\text{D}}}-{{\mathbf{v}}_{\text{D}}} \right\|}_2^{2}}}{2{{\sigma }^{2}}}-\frac{{{\left\| {{{\hat{\mathbf{v}}}}_{\text{A}}}-{{\mathbf{v}}_{\text{A}}} \right\|}_2^{2}}}{2{{\sigma }^{2}}} \right), \\ 
	\end{aligned}
\end{equation}
where (a) holds because ${{\hat{\mathbf{v}}}_{\text{D}}}$ and ${{\hat{\mathbf{v}}}_{\text{A}}}$ are independent. Therefore, maximizing $P\left( \hat{\mathbf{v}}_{\text{D}}, \hat{\mathbf{v}}_{\text{A}} \mid \mathbf{v}_{\text{D}}, \mathbf{v}_{\text{A}} \right)$ is equivalent to minimizing ${\left\| \hat{\mathbf{v}}_{\text{D}} - \mathbf{v}_{\text{D}} \right\|}_2^{2} + {\left\| \hat{\mathbf{v}}_{\text{A}} - \mathbf{v}_{\text{A}} \right\|}_2^{2}$. 

{This problem can be handled by traditional methods such as LS and LMMSE in \cite{ref_LS}. However, both approaches suffer from notable limitations. Specifically, LS and LMMSE ignore the inherent geometric constraints between the direction vectors, leading to degraded estimation performance. In addition, the LMMSE method requires prior knowledge or estimation of the covariance matrix, which is often unavailable or inaccurate in practice, thereby further limiting its applicability. To overcome these limitations, we incorporate intrinsic constraints into the estimation formulation.}

The direction vectors ${{\mathbf{v}}_{\text{D}}}$ and ${{\mathbf{v}}_{\text{A}}}$ are further subject to two constraints. First, they must lie in the same plane as the line connecting the BS and the user, that is,
\begin{equation}\label{equ1.16}
{{\left( {{\mathbf{v}}_{\text{D}}}\times {{\mathbf{v}}_{\text{A}}} \right)}^{\text{T}}}{{\mathbf{v}}^{\text{u}\to \text{b}}}=0,
\end{equation}
where $\mathbf{v}^{\text{u} \to \text{b}} = \mathbf{p}^{\text{b}} - \mathbf{p}^{\text{u}}$. Second, both ${{\mathbf{v}}_{\text{D}}}$ and ${{\mathbf{v}}_{\text{A}}}$ are constrained to be unit-norm vectors.

In summary, the maximum likelihood estimation problem for $\mathbf{v}_{\text{D}}$ and $\mathbf{v}_{\text{A}}$ can be formulated as  
\begin{equation}\label{equ1.17}
\begin{aligned}
   \left( \text{P3} \right)\quad 
   &\underset{{{\mathbf{v}}_{\text{D}}},{{\mathbf{v}}_{\text{A}}}}{\mathop{\min }}\,\quad \left\| {{{\mathbf{\hat{v}}}}_{\text{D}}}-{{\mathbf{v}}_{\text{D}}} \right\|_{2}^{2}+\left\| {{{\mathbf{\hat{v}}}}_{\text{A}}}-{{\mathbf{v}}_{\text{A}}} \right\|_{2}^{2} \\ 
 & \text{s}\text{. t}\text{.}\quad \text{C1:\,\,}{{\left( {{\mathbf{v}}_{\text{D}}}\times {{\mathbf{v}}_{\text{A}}} \right)}^{\text{T}}}{{\mathbf{v}}^{\text{u}\to \text{b}}}=0 \\ 
 & \quad\quad\,\,\, \text{C2:\,\,}{{\left\| {{\mathbf{v}}_{\text{D}}} \right\|}_{2}}={{\left\| {{\mathbf{v}}_{\text{A}}} \right\|}_{2}}=1. \\ 
\end{aligned}
\end{equation}

It is evident that due to the presence of the two constraints, $\mathbf{v}_{\text{D}}$ and $\mathbf{v}_{\text{A}}$ are coupled, which renders the optimization problem non-convex and difficult to solve directly. To address this challenge, a penalty function is introduced to handle constraint (C1). Then, the BCD is employed to alternately optimize $\mathbf{v}_{\text{D}}$ and $\mathbf{v}_{\text{A}}$, while the RMO is used to enforce constraint (C2).

To handle constraint (C1), a penalty function is introduced as  
\begin{equation}\label{equ1.18}
\phi \left( {{\mathbf{v}}_{\text{D}}},{{\mathbf{v}}_{\text{A}}} \right)=\frac{\lambda }{2}{{\left[ {{\left( {{\mathbf{v}}_{\text{D}}}\times {{\mathbf{v}}_{\text{A}}} \right)}^{\text{T}}}{{\mathbf{v}}^{\text{u}\to \text{b}}} \right]}^{2}},
\end{equation}
where $\lambda > 0$ is a weighting factor that controls the strength of the penalty term. With the penalty function incorporated, the optimization problem is reformulated as  
\begin{equation}\label{equ1.19}
\begin{aligned}
   \left( \text{P4} \right)\quad &\underset{{{\mathbf{v}}_{\text{D}}},{{\mathbf{v}}_{\text{A}}}}{\mathop{\min }}\,\quad \mathcal{L}\left( {{\mathbf{v}}_{\text{D}}},{{\mathbf{v}}_{\text{A}}} \right) \\ 
 & \text{s}\text{. t}\text{.}\quad \text{C2:\,\,}{{\left\| {{\mathbf{v}}_{\text{D}}} \right\|}_{2}}={{\left\| {{\mathbf{v}}_{\text{A}}} \right\|}_{2}}=1, \\ 
\end{aligned}
\end{equation}
where  
\begin{equation}\label{equ1.20}
	\mathcal{L}\left( {{\mathbf{v}}_{\text{D}}},{{\mathbf{v}}_{\text{A}}} \right)={{\left\| {{{\hat{\mathbf{v}}}}_{\text{D}}}-{{\mathbf{v}}_{\text{D}}} \right\|}_2^{2}}+{{\left\| {{{\hat{\mathbf{v}}}}_{\text{A}}}-{{\mathbf{v}}_{\text{A}}} \right\|}_2^{2}}+\phi \left( {{\mathbf{v}}_{\text{D}}},{{\mathbf{v}}_{\text{A}}} \right).
\end{equation}

By examining \hyperref[equ1.19]{(P4)}, it can be observed that $\mathcal{L}\left( \mathbf{v}_{\text{D}}, \mathbf{v}_{\text{A}} \right)$ is a multi-convex function with respect to $\mathbf{v}_{\text{D}}$ and $\mathbf{v}_{\text{A}}$ and it can be handled using BCD~\cite{ref1_BCD}.

Problem \hyperref[equ1.19]{(P4)} can be decomposed into two subproblems, where $\mathbf{v}_{\text{D}}$ and $\mathbf{v}_{\text{A}}$ are alternately optimized. Taking the subproblem of optimizing $\mathbf{v}_{\text{D}}$ with fixed $\mathbf{v}_{\text{A}}$ as an example, it can be formulated as  
\begin{equation}\label{equ1.21}
\begin{aligned}
 \left( \text{P4-1} \right)\quad &\underset{{{\mathbf{v}}_{\text{D}}}}{\mathop{\min }}\,\quad \mathcal{L}\left( {{\mathbf{v}}_{\text{D}}} \right) \\ 
 & \text{s}\text{. t}\text{.}\quad \text{C2-1:\,\,}{{\left\| {{\mathbf{v}}_{\text{D}}} \right\|}_{2}}=1,  \\ 
\end{aligned}
\end{equation}
where $\mathcal{L}\left( {{\mathbf{v}}_{\text{D}}} \right)={{\left\| {{{\mathbf{\hat{v}}}}_{\text{D}}}-{{\mathbf{v}}_{\text{D}}} \right\|}_2^{2}}+{\lambda {{({{\mathbf{u}}^{\text{T}}}{{\mathbf{v}}_{\text{D}}})}^{2}}}/{2}$.

The main challenge in solving \hyperref[equ1.21]{(P4-1)} arises from constraint (C2-1), which can be addressed using RMO~\cite{ref2_BCD}. Specifically, at the $t$-th iteration, let the solution from the previous iteration be $\mathbf{v}_{\text{D}}^{\left( t-1 \right)}$. First, the Euclidean gradient of the objective function $\mathcal{L}\left( \mathbf{v}_{\text{D}} \right)$ is computed, denoted as $\mathbf{g}_{{\mathbf{v}}_{\text{D}}}^{\left( t \right)}$, given by  
\begin{equation}\label{equ1.22}
	\mathbf{g}_{{{\mathbf{v}}_{\text{D}}}}^{\left( t \right)}=-{{\nabla }_{{{\mathbf{v}}_{\text{D}}}}}{{\mathcal{L}}^{\left( t \right)}}=2\left( {{{\hat{\mathbf{v}}}}_{\text{D}}}-\mathbf{v}_{\text{D}}^{\left( t \right)} \right)-\lambda \mathbf{u}{{\mathbf{u}}^{\text{T}}}\mathbf{v}_{\text{D}}^{\left( t \right)},
\end{equation}
Then, the Euclidean gradient $\mathbf{g}_{{\mathbf{v}}_{\text{D}}}^{\left( t \right)}$ is projected onto the tangent space of the unit-norm manifold to obtain the Riemannian gradient  
\begin{equation}\label{equ1.23}
	\bar{\mathbf{g}}_{{{\mathbf{v}}_{\text{D}}}}^{\left( t \right)}=\mathbf{g}_{{{\mathbf{v}}_{\text{D}}}}^{\left( t \right)}-\left[ {{\left( \mathbf{g}_{{{\mathbf{v}}_{\text{D}}}}^{\left( t \right)} \right)}^{\text{T}}}\mathbf{v}_{\text{D}}^{\left( t \right)} \right]\mathbf{v}_{\text{D}}^{\left( t \right)}.
\end{equation}
Accordingly, the solution at the $t$-th iteration is updated as  
\begin{equation}\label{equ1.24}
	\bar{\mathbf{v}}_{\text{D}}^{\left( t \right)}=\mathbf{v}_{\text{D}}^{\left( t \right)}+\beta _{\text{D}}^{\left( t \right)}\bar{\mathbf{g}}_{{{\mathbf{v}}_{\text{D}}}}^{\left( t \right)},
\end{equation}
where $\beta_{\text{D}}$ denotes the step size, which can be adaptively updated using the Armijo rule~\cite{ref3_Armijo}. Specifically, when the condition ${{\mathcal{L}}^{\left( t+1 \right)}}\left( {{\mathbf{v}}_{\text{D}}} \right)>{{\mathcal{L}}^{\left( t \right)}}\left( {{\mathbf{v}}_{\text{D}}} \right)+c\beta _{\text{D}}^{\left( t \right)}\bar{\mathbf{g}}_{{{\mathbf{v}}_{\text{D}}}}^{\left( t \right)}$ is satisfied, the step size is updated as $\beta_{\text{D}}^{\left( t+1 \right)} = c \beta_{\text{D}}^{\left( t \right)}$, where $c\in (0,{1}/{2})$. This condition is used to ensure the superlinear convergence of the algorithm.

To ensure that the solution has unit norm, a retraction operation is applied to $\bar{\mathbf{v}}_{\text{D}}^{\left( t \right)}$. The final update result is then given by  
\begin{equation}\label{equ1.25}
	\mathbf{v}_{\text{D}}^{\left( t+1 \right)}\leftarrow \bar{\mathbf{v}}_{\text{D}}^{\left( t \right)}\frac{1}{{{\left\| \bar{\mathbf{v}}_{\text{D}}^{\left( t \right)} \right\|}_{2}}}.
\end{equation}

The subproblem of optimizing $\mathbf{v}_{\text{A}}$ with $\mathbf{v}_{\text{D}}$ fixed can be formulated as  
\begin{equation}\label{equ1.26}
\begin{aligned}
\left( \text{P4-2} \right)\quad &\underset{{{\mathbf{v}}_{\text{A}}}}{\mathop{\min }}\,\quad \mathcal{L}\left( {{\mathbf{v}}_{\text{A}}} \right) \\ 
 & \text{s}\text{. t}\text{.}\quad \text{C2-2:\,\,}{{\left\| {{\mathbf{v}}_{\text{A}}} \right\|}_{2}}=1, \\ 
\end{aligned}
\end{equation}
where $\mathcal{L}\left( {{\mathbf{v}}_{\text{A}}} \right)={{\left\| {{{\mathbf{\hat{v}}}}_{\text{A}}}-{{\mathbf{v}}_{\text{A}}} \right\|}_2^{2}}+{\lambda {{({{\mathbf{w}}^{\text{T}}}{{\mathbf{v}}_{\text{A}}})}^{2}}}/{2}$, where $\mathbf{w}^{\text{T}} = \mathbf{v}^{\text{u} \to \text{b}} \times \mathbf{v}_{\text{D}}$. The solution  is similar to that of \hyperref[equ1.21]{(P4-1)} and is omitted here for brevity.

The detailed steps for the initial RP estimation based on BCD-RMO are summarized in Algorithm~\ref{RM-BCD}, where $T_{\max}$ denotes the maximum number of iterations, $\beta_{\text{A}}$ is the step size used for optimizing $\mathbf{v}_{\text{A}}$, and $\varepsilon_{\mathcal{L}}$ is the convergence threshold. The iteration terminates when $\left| \mathcal{L}^{\left( t+1 \right)} - \mathcal{L}^{\left( t \right)} \right| < \varepsilon_{\mathcal{L}}$.

\begin{algorithm}[htb]
\caption{Initial RP Estimation Based on BCD-RMO}\label{RM-BCD}
\setstretch{1.2}
\begin{algorithmic}
	\Statex{\textbf{Input:} $\hat{\mathbf{v}}_{\text{A}}, \hat{\mathbf{v}}_{\text{D}}, \mathbf{p}^{\text{u}}, \mathbf{p}^{\text{b}}, \sigma, T_{\max}, c, \beta_{\text{D}}, \beta_{\text{A}}, \varepsilon_{\mathcal{L}}$}
	\State 1. \textbf{Initialization:} $\mathbf{v}_{\text{D}}^{(0)} = \hat{\mathbf{v}}_{\text{D}}$, $\mathbf{v}_{\text{A}}^{(0)} = \hat{\mathbf{v}}_{\text{A}}$, set $\beta_{\text{D}}^{(0)}$, $\beta_{\text{A}}^{(0)}$, $t = 0$
	\State 2. \textbf{for} $t = 0 : T_{\max}$ \textbf{do}
	\State \hspace{0.4cm}2.1. \textbf{Fix} $\mathbf{v}_{\text{A}}^{(t)}$ and \textbf{update} $\mathbf{v}_{\text{D}}^{(t+1)}$:
	\State \hspace{0.8cm}2.1.1. Compute $\mathbf{g}_{\mathbf{v}_{\text{D}}}^{(t)}$ of $\mathcal{L}(\mathbf{v}_{\text{D}})$ according to~(\ref{equ1.22});
	\State \hspace{0.8cm}2.1.2. Compute $\bar{\mathbf{g}}_{\mathbf{v}_{\text{D}}}^{(t)}$ according to~(\ref{equ1.23});
	\State \hspace{0.8cm}2.1.3. \textbf{if} $\mathcal{L}^{(t+1)}(\mathbf{v}_{\text{D}}) > \mathcal{L}^{(t)}(\mathbf{v}_{\text{D}}) + c \beta_{\text{D}}^{(t)} \bar{\mathbf{g}}_{\mathbf{v}_{\text{D}}}^{(t)}$
	\State \hspace{2.1cm}$\beta_{\text{D}}^{(t+1)} = c \beta_{\text{D}}^{(t)}$
	\State \hspace{1.7cm}\textbf{else}
	\State \hspace{2.1cm}$\beta_{\text{D}}^{(t+1)} = \beta_{\text{D}}^{(t)}$
	\State \hspace{1.7cm}\textbf{end if}
	\State \hspace{0.8cm}2.1.4. Update $\bar{\mathbf{v}}_{\text{D}}^{(t)}$ according to~(\ref{equ1.24});
	\State \hspace{0.8cm}2.1.5. Normalize to obtain $\mathbf{v}_{\text{D}}^{(t+1)}$ according to~(\ref{equ1.25});
	\State \hspace{0.4cm}2.2. \textbf{Fix} $\mathbf{v}_{\text{D}}^{(t+1)}$ and \textbf{update} $\mathbf{v}_{\text{A}}^{(t+1)}$:
	\State \hspace{1.2cm}Same as step 2.1
	\State \hspace{0.4cm}2.3. \textbf{if} $\left| \mathcal{L}^{(t+1)} - \mathcal{L}^{(t)} \right| < \varepsilon_{\mathcal{L}}$, \textbf{break}
	\State \hspace{1cm}\textbf{end if}
	\State \hspace{0.4cm}\textbf{end for}
	\State 3. \textbf{Output:} $\mathbf{v}_{\text{A}}^*$, $\mathbf{v}_{\text{D}}^*$
	\State 4. Compute the final RP position $\mathbf{p}^{\text{r}^*}$ according to~(\ref{equ1.10})--(\ref{equ1.14}).
\end{algorithmic}
\end{algorithm}

\subsection{Stage 2: RP Refinement Based on VBS-HC}
{ \color{black}The first-stage algorithm estimates the RP position corresponding to each signal path. As shown in Fig.~\ref{fig3.3}(a), due to random noise, the estimated RP of a single path may still exhibit a large deviation. To further enhance estimation accuracy and robustness, Stage 2 introduces the VBS to refine RP positions.  Multiple paths in the same SA are reflected by the same object and should correspond to a unique VBS \cite{rel2_SLAM}, while they yield different RP positions. Therefore, the VBS positions for each path are inferred first and then  clustered for SA information aggregation. The centroid of each VBS cluster is used to refine the RP positions.}

\begin{figure}[!thb]
	\centering
        {\includegraphics[width=\linewidth]{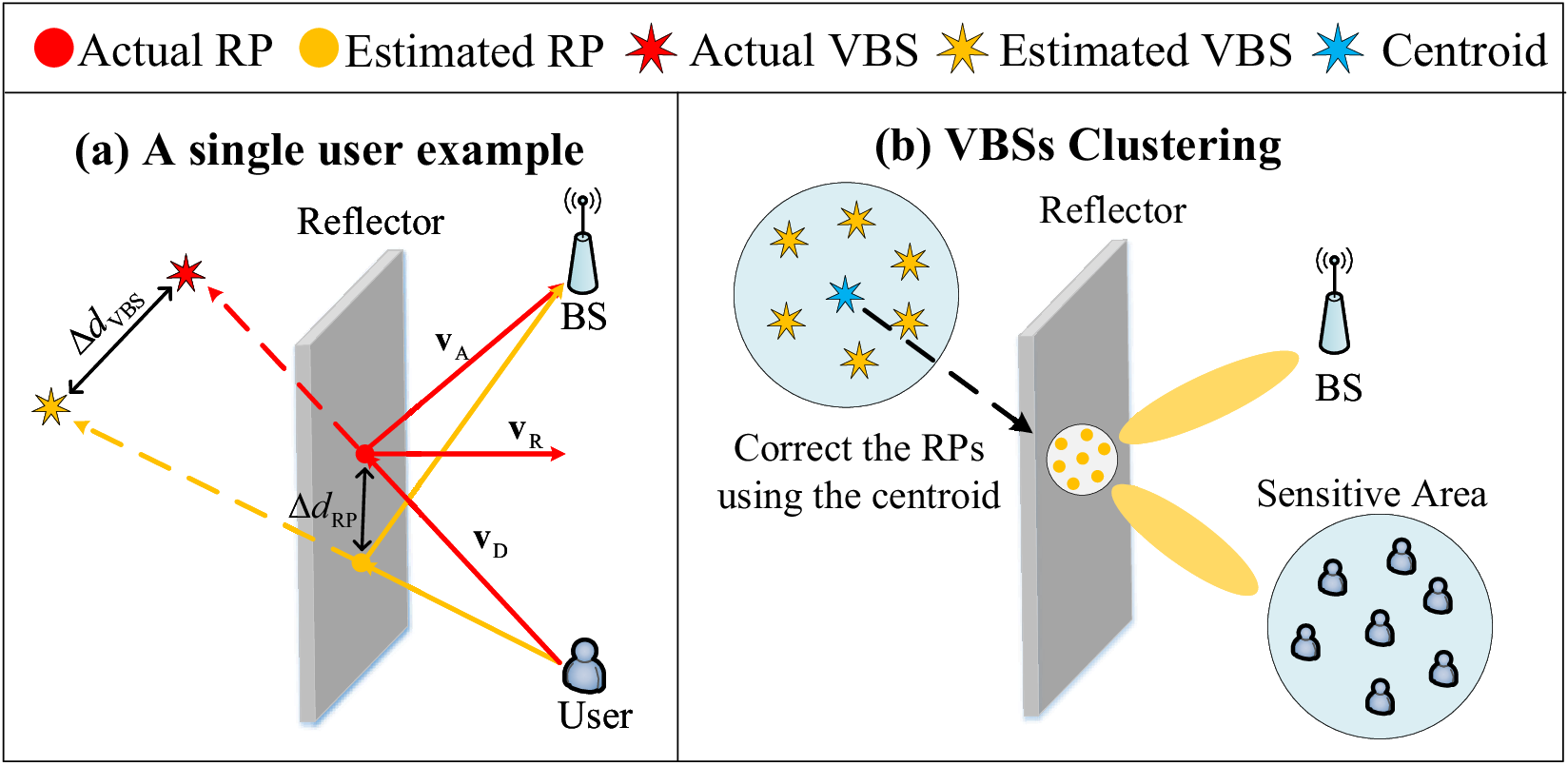}}
	\caption{Illustration of the VBS clustering enhanced RP position estimation. (a) shows the deviation between the ground-truth  and first-stage estimated positions of the RP and VBS for a single user, where $\mathbf{v}_\text{R}$ is the normal vector of the reflector surface. (b) depicts the VBS clustering in the multi-user case, where VBS positions derived from different users are clustered, and the centroid is used to improve the RP position estimation.}
	\label{fig3.3}
\end{figure}

{\color{black}Given ${\mathbf p}^{\text b}$, ${\mathbf p}^{\text u}_{k}$ and the estimated RP position in Stage 1, the propagation distance ${{d}_{l}}$ of path $l$ can be easily obtained. In practice, ${{d}_{l}}$ can also be derived from the propagation delay~\cite{ref_gqd}. In this case, the VBS position corresponding to the reference path $l$ can be expressed as}
\begin{equation}\label{equ1.27}
	\mathbf{p}_{l}^{\text{v}}={{\mathbf{p}}^{\text{u}}}+{{d}_{l}}{{\mathbf{v}}_{\text{D}}}.
\end{equation}

Assuming that the first-stage algorithm estimates the RPs corresponding to $L=\sum\nolimits_{k=1}^{K}{{{L}_{k}}}$ paths, let the corresponding VBS position set be denoted as $\mathcal{P}=\left\{ \mathbf{p}_{1}^{\text{v}},\cdots ,\mathbf{p}_{L}^{\text{v}} \right\}$. In this work, the  AGglomerative NESting (AGNES) algorithm~\cite{ref_gqd} is adopted to perform hierarchical clustering on $\mathcal{P}$, resulting in a centroid set $\{\mathbf{p}_{1}^{\text{g}},\cdots ,\mathbf{p}_{m}^{\text{g}},\cdots ,\mathbf{p}_{{{M}_{1}}}^{\text{g}}\}$ that represents the VBS positions corresponding to each reflector.

Using the centroids obtained from VBS clustering, the RP position of each path can be corrected. As illustrated in Fig.~\ref{fig3.3}(a), suppose the centroid of the VBS cluster corresponding to a reference path is denoted as $\mathbf{p}_{m}^{\text{g}}$, which is then taken as the updated VBS position for this path. The departure direction vector ${{\mathbf{v}}_{\text{D}}}$ is updated accordingly as follows
\begin{equation}\label{equ1.28}
	{{\mathbf{v}}_{\text{D}}}=\frac{\mathbf{p}_{m}^{\text{g}}-\mathbf{p}_{k}^{\text{u}}}{\left\| \mathbf{p}_{m}^{\text{g}}-\mathbf{p}_{k}^{\text{u}} \right\|}_2,
\end{equation}
Based on geometric relations, the normal vector of the reflecting surface ${{\mathbf{v}}_{\text{R}}}$ can be obtained as
\begin{equation}\label{equ1.29}
	{{\mathbf{v}}_{\text{R}}}=\frac{{{\mathbf{p}}^{\text{b}}}-\mathbf{p}_{m}^{\text{g}}}{\left\| {{\mathbf{p}}^{\text{b}}}-\mathbf{p}_{m}^{\text{g}} \right\|}_2,
\end{equation}
The updated ${{\mathbf{v}}_{\text{A}}}$ can be further computed as
\begin{equation}\label{equ1.30}
	{{\mathbf{v}}_{\text{A}}}={{\mathbf{v}}_{\text{D}}}-2\left( {{\mathbf{v}}_{\text{R}}}\mathbf{v}_{\text{D}}^{\text{T}} \right){{\mathbf{v}}_{\text{R}}}.
\end{equation}

To further improve the robustness of the VBS estimation, weighting factors ${{\alpha }_{\text{D}}}$ and ${{\alpha }_{\text{A}}}$ are introduced to fuse the updated ${{\mathbf{v}}_{\text{D}}}$ and ${{\mathbf{v}}_{\text{A}}}$ with their initial observations ${{\hat{\mathbf{v}}}_{\text{D}}}$ and ${{\hat{\mathbf{v}}}_{\text{A}}}$, respectively
\begin{equation}\label{equ1.31}
	\begin{aligned}
		& {{\mathbf{v}}_{\text{D}}}\leftarrow \alpha_{\text{D}} {{\mathbf{v}}_{\text{D}}}+\left( 1-\alpha_{\text{D}}  \right){{{\hat{\mathbf{v}}}}_{\text{D}}}, \\ 
		& {{\mathbf{v}}_{\text{A}}}\leftarrow \alpha_{\text{A}} {{\mathbf{v}}_{\text{A}}}+\left( 1-\alpha_{\text{A}}  \right){{{\hat{\mathbf{v}}}}_{\text{A}}}. \\ 
	\end{aligned}
\end{equation}
The introduction of weighting factors ${{\alpha }_{\text{D}}}$ and ${{\alpha }_{\text{A}}}$ balances the relationship between the historical iterative values and the original observations, effectively mitigating the impact of single-shot observation noise.

{\color{black}The aforementioned second-stage algorithm is referred to as the RP Refinement based on VBS Hierarchical Clustering (VBS-HC). The detailed procedure is shown in Algorithm~\ref{VBS}, where ${{N}_{\max }}$ denotes the maximum number of iterations and ${{\varepsilon }_{\text{v}}}$ is the convergence threshold.
Based on the refined ${{\mathbf{v}}_{\text{D}}}$ and ${{\mathbf{v}}_{\text{A}}}$, the first-stage algorithm is re-executed to determine the RP position for the corresponding path.} 
\begin{algorithm}[htb]
\caption{RP Refinement Algorithm Based on VBS-HC}\label{VBS}
\setstretch{1.2}
	\begin{algorithmic}
	\Statex{\textbf{Input:} ${{\mathbf{p}}^{\text{u}}}$, ${{\mathbf{p}}^{\text{b}}}$, $\mathcal{P}$, ${{N}_{\max }}$, ${L}$, ${{\varepsilon }_{\text{v}}}$, $\alpha_{\text{D}}$, $\alpha_{\text{A}}$}
	\State 1: \textbf{Initialization:} ${{\mathcal{P}}^{\left( 0 \right)}}$, $n=0$
	\State 2: \textbf{for} $n = 0 : {{N}_{\max }}$ \textbf{do}
	\State\hspace{0.4cm}2.1: Perform hierarchical clustering on ${{\mathcal{P}}^{\left( n \right)}}$ to obtain the centroid set $\left\{ \mathbf{p}_{1}^{\text{g,}(n)}, \cdots, \mathbf{p}_{m}^{\text{g,}(n)}, \cdots, \mathbf{p}_{{{M}_{1}}}^{\text{g,}(n)} \right\}$
	\State\hspace{0.4cm}2.2: \textbf{for} $l = 1 : {L}$ \textbf{do}
	
	\State\hspace{0.8cm}2.2.1: Assign centroid $\mathbf{p}_{m}^{\text{g,}(n)}$ to the VBS position of path $l$: $\mathbf{p}_{l}^{\text{v,}(n)} \leftarrow \mathbf{p}_{m}^{\text{g,}(n)}$;
	\State\hspace{0.8cm}2.2.2: Update $\mathbf{v}_{\text{D},l}^{(n)}$, $\mathbf{v}_{\text{R},l}^{(n)}$, and $\mathbf{v}_{\text{A},l}^{(n)}$ using (\ref{equ1.28})–(\ref{equ1.30});
	\State\hspace{0.8cm}2.2.3: Apply fusion weights to $\mathbf{v}_{\text{D},l}^{(n)}$ and $\mathbf{v}_{\text{A},l}^{(n)}$ according to (\ref{equ1.31});
	\State\hspace{0.8cm}2.2.4: Compute $\mathbf{p}_{l}^{\text{r,}(n)}$ using (\ref{equ1.10})--(\ref{equ1.14});
	\State\hspace{0.8cm}2.2.5: Update $\mathbf{p}_{l}^{\text{v,}(n)}$ using (\ref{equ1.27});
	\State\hspace{0.8cm}\textbf{end for}
	\State\hspace{0.4cm}2.3: Update ${{\mathcal{P}}^{(n+1)}}$;
	\State\hspace{0.4cm}2.4: \textbf{if} $\frac{1}{L} \sum\limits_{l=1}^{L} \left\| \mathbf{p}_{l}^{\text{v,}(n+1)} - \mathbf{p}_{l}^{\text{v,}(n)} \right\|_2 \le {{\varepsilon }_{\text{v}}}$, \textbf{break}
	\State\hspace{1.1cm}\textbf{end if}
	\State\hspace{0.6cm}\textbf{end for}
	\State 3: \textbf{Output:} Refined ${{\mathbf{p}}^{{{\text{r}}^{*}}}}$, ${{\mathcal{P}}^{*}}$
	\end{algorithmic}
\end{algorithm}

The two-stage algorithm iteratively refines the RP positions until the desired accuracy is achieved. In the first stage, the BCD-RMO algorithm performs initial estimation of the RPs based on single-path observation information. In the second stage, the VBS positions corresponding to multiple paths within the same SA are clustered and fused to refine the RP estimates of each path. This method is referred to in this paper as the \underline{I}nformation \underline{A}ggregation Based Robust \underline{T}wo-\underline{S}tage \underline{R}eflection \underline{P}oint \underline{R}efinement algorithm (IA-TS-RPR).

\section{Environment Variation Sensing Based on the HNN Criterion and Time-Window Voting}
In Section \Rmnum{4}, the RP positions and their corresponding VBSs for each path were obtained. It is also noted that in the same SA, different paths share a unique VBS position associated with a specific reflector. In this work, we will leverage this property to realize environment variation sensing.

As indicated by \hyperref[equ1.8]{(P1)}, the estimation of $\Delta {E}$ is dependent on the comparison between ${{\mathcal{R}}_{\text{C}}}$ and ${{\mathcal{R}}_{\text{U}}}$. By employing the IA-TS-RPR algorithm to process the path information at each location within ${{\mathcal{R}}_{\text{C}}}$ and ${{\mathcal{R}}_{\text{U}}}$, the corresponding sets of VBSs can be derived, denoted respectively as $\{\mathcal{P}_{1}^{\text{c}},\ldots ,\mathcal{P}_{{{N}_{\text{CKM}}}}^{\text{c}}\}$ and $\{\mathcal{P}_{1}^{\text{u}},\ldots ,\mathcal{P}_{K}^{\text{u}}\}$. Let
\begin{equation}\label{equ1.32}
	{{\mathcal{Q}}_{\text{C}}}=\left\{ \left\langle \mathbf{p}_{1}^{\text{c}},\mathcal{P}_{1}^{\text{c}} \right\rangle ;\left\langle \mathbf{p}_{2}^{\text{c}},\mathcal{P}_{2}^{\text{c}} \right\rangle ;\ldots ;\left\langle \mathbf{p}_{{{N}_{\text{CKM}}}}^{\text{c}},\mathcal{P}_{{{N}_{\text{CKM}}}}^{\text{c}} \right\rangle  \right\},
\end{equation}
which represents the set of VBS distribution features associated with the CKM, which includes the GP location information in the CKM as well as the corresponding set of VBS positions. Similarly, let 
\begin{equation}\label{equ1.33}
	{{\mathcal{Q}}_{\text{U}}}=\left\{ \left\langle \mathbf{p}_{1}^{\text{u}},\mathcal{P}_{1}^{\text{u}} \right\rangle ;\left\langle \mathbf{p}_{2}^{\text{u}},\mathcal{P}_{2}^{\text{u}} \right\rangle ;\ldots ;\left\langle \mathbf{p}_{K}^{\text{u}},\mathcal{P}_{K}^{\text{u}} \right\rangle  \right\},
\end{equation}
which represents the set of VBS distribution features constructed based on users' real-time channel information, which includes the positions of $K$ users and their corresponding VBS location information.  

Let user $k$ be located at the $\text{ }\!\!\pi\!\!\text{ }(k)$-th GP of the CKM, i.e., $\mathbf{p}_{k}^{\text{u}}=\mathbf{p}_{\text{ }\!\!\pi\!\!\text{ }\left( k \right)}^{\text{c}}$. Then, the difference between $\mathcal{P}_{\text{ }\!\!\pi\!\!\text{ }\left( k \right)}^{\text{c}}$ and $\mathcal{P}_{k}^{\text{u}}$ reflects the variation of reflectors in the environment. Therefore, by utilizing ${{\mathcal{Q}}_{\text{C}}}$ and ${{\mathcal{Q}}_{\text{U}}}$, (P1) can be reformulated as  
\begin{equation}\label{equ1.34}
	\left( \text{P1-T} \right)\quad \widehat{\Delta {E}}=\arg \underset{\Delta {E}}{\mathop{\max }}\,\quad P\left( \Delta {E}\left| {{\mathcal{Q}}_{\text{C}}},{{\mathcal{Q}}_{\text{U}}},{{E}^0} \right. \right).
\end{equation}
The solution to \hyperref[equ1.34]{(P1-T)} is challenging. First, the actual user position $\mathbf{p}_{k}^{\text{u}}$ may not coincide with any CKM GP. Therefore, it is necessary to identify the CKM GP $\mathbf{p}_{\text{ }\!\!\pi\!\!\text{ }\left( k \right)}^{\text{c}}$ that exhibits the strongest association with the channel information at $\mathbf{p}_{k}^{\text{u}}$, i.e., to properly design $\text{ }\!\!\pi\!\!\text{ }(k)$. Second, the presence of multiple reflectors in the environment may lead to discrepancies in the number of channel paths between nearby positions. As a result, the difference between $\mathcal{P}_{\text{ }\!\!\pi\!\!\text{ }\left( k \right)}^{\text{c}}$ and $\mathcal{P}_{k}^{\text{u}}$ may arise from positional deviations rather than actual variations in the reflectors, which can easily result in misjudgments.

To address the aforementioned challenges, this paper first employs the Hausdorff distance to quantify the difference between $\mathcal{P}_{\text{ }\!\!\pi\!\!\text{ }\left( k \right)}^{\text{c}}$ and $\mathcal{P}_{k}^{\text{u}}$, and proposes a $\text{ }\!\!\pi\!\!\text{ }(k)$ design based on the HNN criterion to establish the correspondence between user locations and CKM GPs. Subsequently, a time-window-based multi-user voting mechanism is introduced to mitigate the uncertainty in judgment caused by positional deviations.

We first introduce the $\text{ }\!\!\pi\!\!\text{ }(k)$ design based on the HNN criterion. As illustrated in Fig.~\ref{fig3.5}, the ${{N}_{\text{H}}}$ nearest GPs surrounding user $k$ are selected, and the corresponding VBS sets are denoted as $\{\mathcal{P}_{{{N}_{1}}}^{\text{c}},\ldots ,\mathcal{P}_{{{N}_{\text{H}}}}^{\text{c}}\}$. Then, the Hausdorff distance between $\mathcal{P}_{{{N}_{\text{h}}}}^{\text{c}}$ (where $1\le {{N}_{\text{h}}}\le {{N}_{\text{H}}}$) and $\mathcal{P}_{k}^{\text{u}}$, denoted as $\text{H}(\mathcal{P}_{{{N}_{\text{h}}}}^{\text{c}},\mathcal{P}_{k}^{\text{u}})$, can be expressed as  
\begin{equation}\label{equ1.35}
	\text{H}\left( \mathcal{P}_{{{N}_{\text{h}}}}^{\text{c}},\mathcal{P}_{k}^{\text{u}} \right)=\max \left\{ \text{h}\left( \mathcal{P}_{{{N}_{\text{h}}}}^{\text{c}},\mathcal{P}_{k}^{\text{u}} \right)\text{,h}\left( \mathcal{P}_{k}^{\text{u}},\mathcal{P}_{{{N}_{\text{h}}}}^{\text{c}} \right) \right\},
\end{equation}
where $\text{h}\left( \mathcal{A},\mathcal{B} \right)=\underset{a\in \mathcal{A}}{\mathop{\max }}\,\underset{b\in \mathcal{B}}{\mathop{\min }}\,\quad {{\left\| a-b \right\|}_{2}}$ denotes the directed distance from set $\mathcal{A}$ to set $\mathcal{B}$. The Hausdorff distance between two sets reflects their degree of similarity. Therefore, a smaller $\text{H}(\mathcal{P}_{{{N}_{\text{h}}}}^{\text{c}},\mathcal{P}_{k}^{\text{u}})$ indicates a higher similarity between the VBS sets corresponding to $\mathbf{p}_{k}^{\text{u}}$ and $\mathbf{p}_{{{N}_{\text{h}}}}^{\text{c}}$, which implies that the associated channel characteristics at the two positions are more alike.

Among the ${{N}_{\text{H}}}$ neighboring GPs, the one with the minimum Hausdorff distance to user $k$ is given by  
\begin{equation}\label{equ1.36}
	\text{ }\!\!\pi\!\!\text{ }(k)=\arg \underset{{{N}_{\text{h}}}}{\mathop{\min }}\,\quad \text{H}\left( \mathcal{P}_{k}^{\text{u}},\mathcal{P}_{{{N}_{\text{h}}}}^{\text{c}} \right),
\end{equation}
Let $d_{k}^{\text{H}}=\text{H}(\mathcal{P}_{k}^{\text{u}},\mathcal{P}_{\pi(k)}^{\text{c}})$ denote the minimum Hausdorff distance corresponding to user $k$. Let the associated VBS position match be $(\mathbf{p}_{\text{H},k}^{\text{v}},\mathbf{p}_{\text{H},\text{ }\!\!\pi\!\!\text{ }\left( k \right)}^{\text{v}})$, where $\mathbf{p}_{\text{H},k}^{\text{v}}\in \mathcal{P}_{k}^{\text{u}}$ and $\mathbf{p}_{\text{H},\text{ }\!\!\pi\!\!\text{ }\left( k \right)}^{\text{v}}\in \mathcal{P}_{\text{ }\!\!\pi\!\!\text{ }\left( k \right)}^{\text{c}}$. This position match is employed for subsequent abnormal path identification and RP location estimation. 
{\color{black}Furthermore, the directed Hausdorff distances in \eqref{equ1.35} can be used to distinguish two types of changes: i) $\text{h}\!\left( \mathcal{P}_k^{\text{u}},\mathcal{P}_{{N_{\text{h}}}}^{\text{c}} \right)$ characterizes the newly emerged VBS (i.e., reflector) that cannot be matched by the CKM; ii)  $\text{h}\!\left( \mathcal{P}_{{N_{\text{h}}}}^{\text{c}},\mathcal{P}_k^{\text{u}} \right)$ characterizes the vanished VBS that can no longer be detected by the users.}

\begin{figure}[!thb]
	\centering
	{\includegraphics[width=0.6\linewidth]{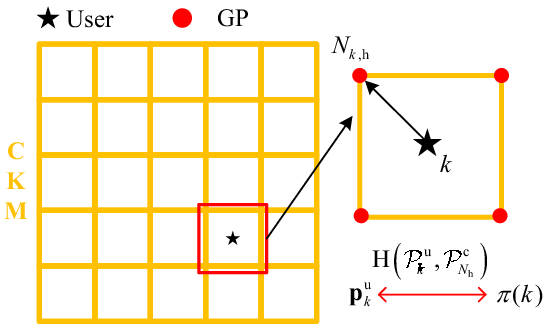}}
	\caption{Illustration of user-to-grid matching based on the CKM. Since user positions do not coincide with CKM GPs, the optimal match $\text{ }\!\!\pi\!\!\text{ }\left( k \right)$ is selected within the user’s neighborhood based on the Hausdorff distance.}
	\label{fig3.5}
\end{figure}

The time-window-based multi-user voting mechanism is introduced as follows. By enabling multiple users to collaboratively determine whether an environmental variation has occurred, the probability of misjudgment can be effectively reduced. However, if users are located far apart, they may fall within different SAs associated with distinct reflectors, as illustrated in Fig.~\ref{fig3.1}. Therefore, in this work, the BS coverage area $\mathcal{C}$ is uniformly partitioned into ${{N}_{\text{GU}}}$ grid units (GUs). Users within each GU jointly vote to determine whether the associated RP has undergone a change.

Consider a sequence of consecutive time slots $\text{T}{{\text{S}}_{1}},\text{T}{{\text{S}}_{2}},\ldots$. Let ${{\mathcal{K}}_{{{n}_{\text{GU}}},m}}$ denote the set of users located within the ${{n}_{\text{GU}}}$-th GU during time slot $\text{T}{{\text{S}}_{m}}$, and denote its cardinality as $|{{\mathcal{K}}_{{{n}_{\text{GU}}},m}}|={{K}_{{{n}_{\text{GU}}},m}}$. Each user ${{k}_{m}}\in {{\mathcal{K}}_{{{n}_{\text{GU}}},m}}$ first determines whether a change in environmental reflectors has occurred by evaluating the minimum Hausdorff distance. The decision result is denoted as ${{D}_{{{k}_{m}}}}$, where
\begin{equation}\label{equ1.37}
	{{D}_{{{k}_{m}}}}=\left\{ \begin{aligned}
		& 1,\quad d_{k}^{\text{H}}>{{\varepsilon }_{\text{H}}} \\ 
		& 0,\quad d_{k}^{\text{H}}\le {{\varepsilon }_{\text{H}}}. \\ 
	\end{aligned} \right.
\end{equation}
where ‘1’ indicates that a change in environmental reflectors has occurred, while ‘0’ indicates no change. The parameter ${{\varepsilon }_{\text{H}}}$ denotes a predefined decision threshold. To ensure the reliability of environmental change detection, let $B$ denote the minimum number of votes required to validate a change. Then, the BS determines whether an environmental variation has occurred in the ${{n}_{\text{GU}}}$-th GU during the ${{M}_{\text{D,}{{n}_{\text{GU}}}}}$-th time slot, given by
\begin{equation}\label{equ1.38}
\begin{aligned}
  {{M}_{\text{D,}{{n}_{\text{GU}}}}}=&\arg \min \quad m \\ 
 &\text{s}\text{. t}\text{.}\quad \sum\limits_{m=1,2,\ldots }{{{K}_{{{n}_{\text{GU}}},m}}}\ge B. \\ 
\end{aligned}
\end{equation}

Let $A=\sum\nolimits_{m=1}^{{{M}_{\text{D,}{{n}_{\text{GU}}}}}}{{{K}_{{{n}_{\text{GU}}},m}}}$. The final decision result for the ${{n}_{\text{GU}}}$-th GU is denoted as ${{D}_{{{n}_{\text{GU}}}}}$, where
\begin{equation}\label{equ1.39}
{{D}_{{{n}_{\text{GU}}}}}=\left\{ \begin{aligned}
  & 1,\quad \quad {\sum\limits_{m=1}^{{{M}_{\text{D,}{{n}_{\text{GU}}}}}}{\sum\limits_{{{k}_{m}}\in {{\mathcal{K}}_{{{n}_{\text{GU}}},m}}}{{{D}_{{{k}_{m}}}}}}}/{A}\;\ge {{\varepsilon }_{D}} \\ 
 & 0,\quad \quad \text{otherwise}, \\ 
\end{aligned} \right.
\end{equation}
where ${{\varepsilon }_{D}}\in \left( 0,1 \right)$ denotes the voting threshold.

{\color{black}To further localize the changed RP, the GU with ${{D}_{{{n}_{\text{GU}}}}}=1$ is identified as an anomalous region. Users with decision results ${{D}_{{{k}_{m}}}}=1$ during the first ${{M}_{\text{D,}{{n}_{\text{GU}}}}}$ time slots are extracted to form the abnormal user set $\mathcal{K}_{{{n}_{\text{GU}}}}^{\left( 1 \right)}$. For each $k\in \mathcal{K}_{{{n}_{\text{GU}}}}^{\left( 1 \right)}$, the VBS position $\mathbf{p}_{\text{H},k}^{\text{v}}$ corresponding to the minimum Hausdorff distance is used to compute the RP location $\mathbf{p}_{\text{H},k}^{\text{r}}$ by the RP Refinement Algorithm Based on VBS-HC in Section \Rmnum{4}-B.}

To summarize, the proposed scheme first employs the HNN criterion for user-to-grid matching, followed by a time-window-based multi-user voting mechanism to sense environmental variation. This scheme is referred to as \underline{E}nvironment \underline{V}ariation \underline{S}ensing based on \underline{H}ausdorff-Nearest Neighbor \underline{C}riterion and \underline{T}ime-\underline{W}indow \underline{V}oting (EVS-HC-TWV).

\section{Numerical Results And Discussion}
This section presents the performance of the proposed environmental change sensing under different simulation scenarios. 
\subsection{Scenario Description}
{\color{black}As illustrated in Fig.~\ref{fig3.7}, a representative scenario of size 60\,m $\times$ 60\,m is considered (Map 0). The BS is located at the horizontal coordinate origin with a height of 15 meters, operating at the 28\,GHz frequency band with a transmit power of 40\,dBm. 
The area is uniformly divided into GPs with a grid spacing of $d_{\mathrm{ckm}}=1\,\mathrm{m}$.}  
The positions of these GPs, along with their corresponding channel information, constitute the CKM. The channel information is generated using the Raytrace algorithm implemented in MATLAB R2024a.

This work focuses on sensing changes in environmental reflectors and designs a variety of simulation scenarios accordingly. As shown in Fig. \ref{fig3.7}, Map 0 contains static buildings and is used for constructing the CKM, whilst Maps~1 to 4 correspond to four distinct environmental change scenarios with randomly introduced reflectors. The BS compares the uplink channel information collected at different user locations with the CKM data to determine the positions of changed reflectors.

\begin{figure}[!thb]
	\centering
	{\includegraphics[width=.8\linewidth]{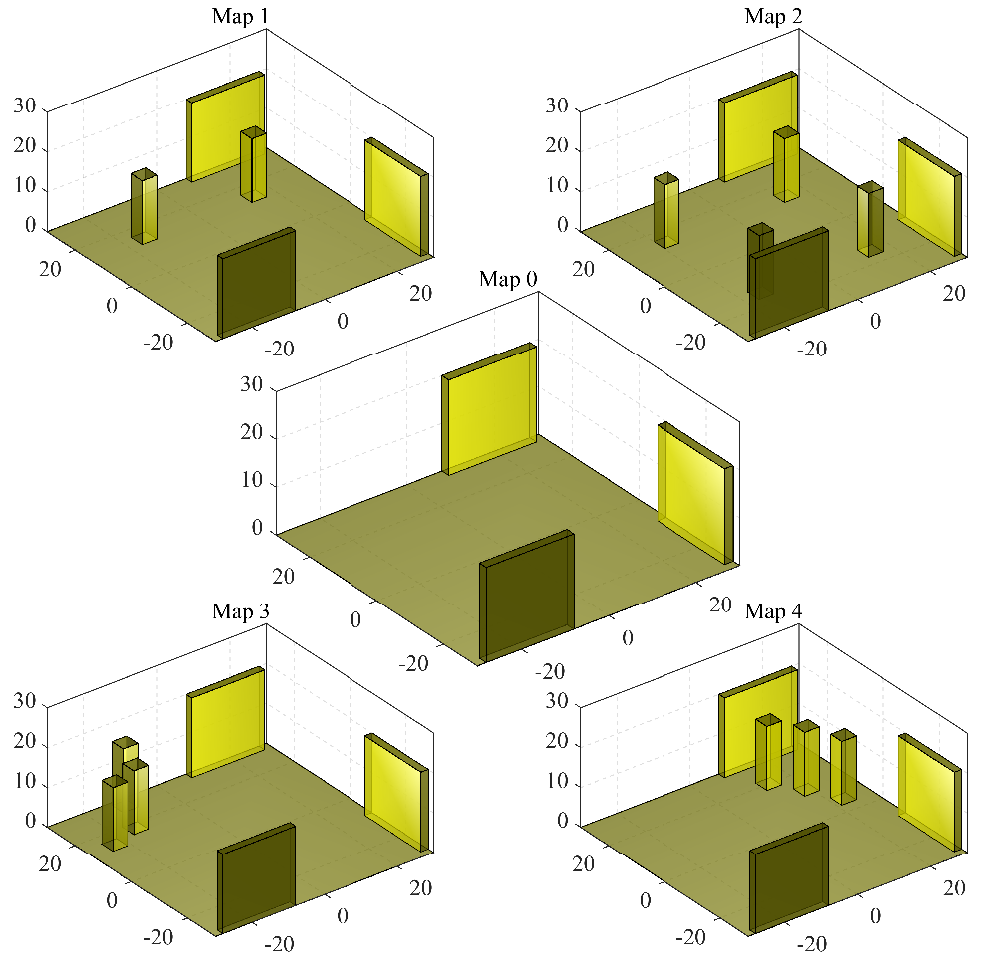}}
	\caption{Illustrations of environment maps, among which Maps 1-4 contain additional randomly placed reflectors compared with the initial environment Map 0.}
	\label{fig3.7}
\end{figure}

\begin{table}[!htb]
	\centering
	\caption{Algorithm parameter settings} \label{table3.3}
	\centering
        \scalebox{1}{
		\begin{tabular}{ll|ll}
			\toprule[1.5pt]
			Parameter & Value & Parameter & Value\\
			\midrule[0.75pt]
			$\beta _{\text{D}}^{\left( 0 \right)}$ & 0.5 & $\beta _{\text{A}}^{\left( 0 \right)}$  & 0.5 \\
			$c$  & 0.5 & $\alpha_{\text{D}}$,$\alpha_{\text{A}}$  & 0.2 \\
			${{\varepsilon }_{\mathcal{L}}}$,${{\varepsilon }_{\text{v}}}$  & ${{10}^{-3}}$ & ${{\varepsilon }_{\text{H}}}$  & 0.3  \\
			${{T}_{\max }}$  & 1000 & ${{N}_{\max }}$  & 100 \\
			\bottomrule[1.5pt]
	\end{tabular}}
\end{table}

\subsection{Performance of RP Position Estimation}
In the simulation scenario shown in Fig.~\ref{fig3.7}, channel data from 500 randomly selected positions are extracted to evaluate the performance of the proposed IA-TS-RPR algorithm. Detailed algorithm parameters are provided in Table~\ref{table3.3}. 
To assess the algorithm's robustness under significant observation errors, this work sets ${{\sigma }^{2}}=3$.
\begin{figure}[!thb]
	\centering
	{\includegraphics[width=.9\linewidth]{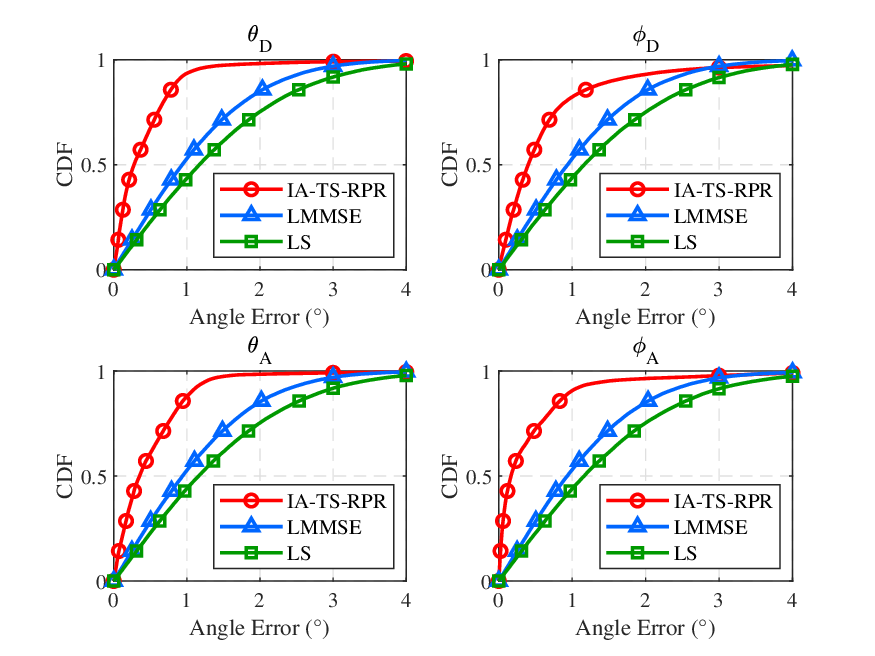}}
	\caption{CDFs of the absolute estimation error for four angles with IA-TS-RPR, LMMSE and LS, respectively.}
	\label{fig3.9}
\end{figure}
\begin{figure}[!thb]
	\centering
	{\includegraphics[width=\linewidth]{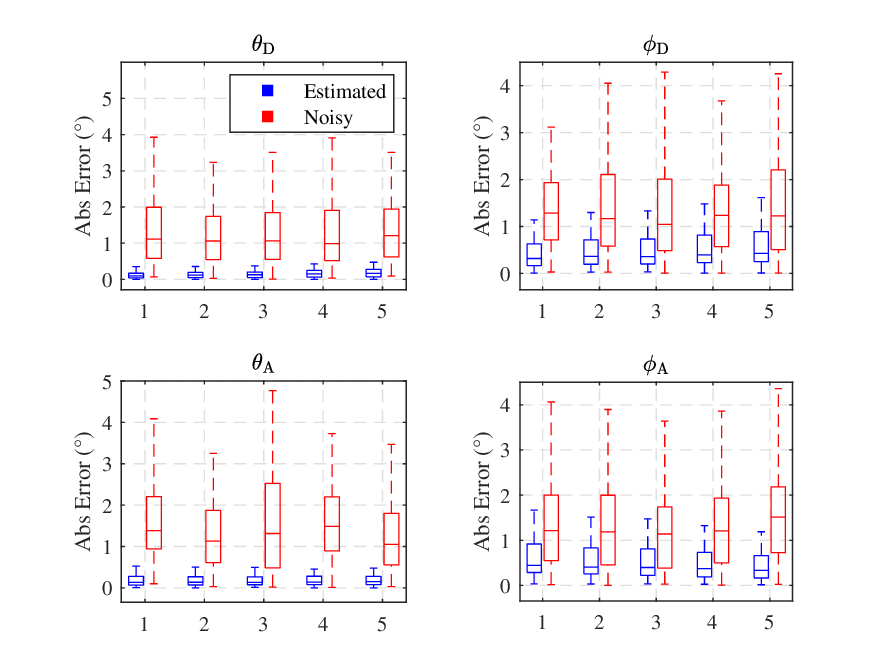}}
	\caption{Box plots of  absolute estimation error for four angles under IA-TS-RPR and raw noise for five randomly selected paths.}
	\label{fig3.8}
\end{figure}

Fig.~\ref{fig3.9} presents the cumulative distribution function (CDF) of the angular estimation errors for the proposed IA-TS-RPR algorithm compared with LS and LMMSE. It can be observed that the error distributions of LS and LMMSE are generally wider across all angular components, with slower CDF growth and 90\% of the errors falling within 2–3°. In contrast, the proposed algorithm exhibits a significant advantage in estimation accuracy across all angular components, with the majority of errors confined within 1°.
Fig.~\ref{fig3.8} shows the box plots of angular estimation errors for five randomly selected paths. As illustrated, the IA-TS-RPR algorithm demonstrates strong noise resilience across all paths. Taking the elevation ${{\theta }_{\text{D}}}$ and ${{\theta }_{\text{A}}}$ as examples, the median error on each path remains below 0.5°, with a relatively narrow spread, indicating high robustness. In comparison, the estimation accuracy of the azimuth ${{\phi }_{\text{D}}}$ and ${{\phi }_{\text{A}}}$ is slightly lower, though the overall error remains within 1°.

As can be seen in  Fig. \ref{fig3.9} and Fig.~\ref{fig3.8}, the estimation errors of azimuth angles are generally higher than those of elevation angles. This phenomenon can be attributed to two main reasons. First, in most practical scenarios, the relative height between the user and the BS is significantly smaller than the horizontal distance, resulting in smaller vertical components in ${{\mathbf{v}}_{\text{D}}}$ and ${{\mathbf{v}}_{\text{A}}}$, which are primarily influenced by the horizontal direction. Second, the IA-TS-RPR algorithm directly estimates ${{\mathbf{v}}_{\text{D}}}$ and ${{\mathbf{v}}_{\text{A}}}$, and the angles are subsequently derived from these vectors. Although the horizontal components of ${{\mathbf{v}}_{\text{D}}}$ and ${{\mathbf{v}}_{\text{A}}}$ are relatively large, the azimuth $\phi$ is computed as the ratio between horizontal components, making it more sensitive to relative perturbations and thus more prone to amplified estimation errors in the azimuth domain.

\begin{figure}[!thb]
	\centering
	{\includegraphics[width=\linewidth]{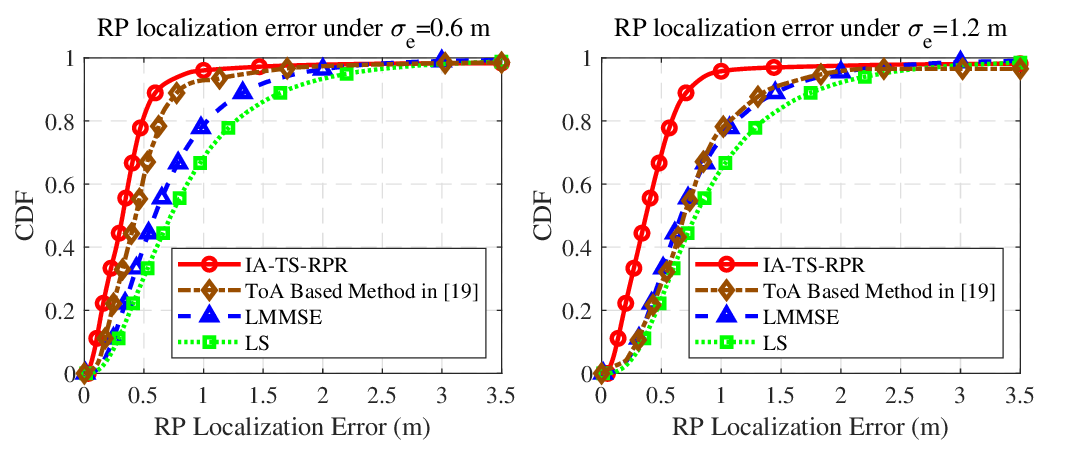}}
	\caption{{\color{black}CDFs of the RP localization error under user positioning errors.}}
	\label{compare_0p6_1p2_2}
\end{figure}

{\color{black}In current ISAC systems, the achievable user localization accuracy is typically at the sub-meter to meter level \cite{user_error1,user_error2}. To verify the robustness of the proposed method against such practical positioning uncertainty, we further introduce user positioning errors that follow $\mathcal{N}(0,\sigma_{\text e}^2)$ with two different uncertainty levels, i.e., $\sigma_{\text e}=0.6$~m and $1.2$~m. The proposed scheme is compared with existing methods including the ToA based method in \cite{Xie2024ToA}, LS and LMMSE. As shown in Fig.~\ref{compare_0p6_1p2_2}, the proposed IA-TS-RPR achieves smaller RP localization errors than all its counterparts in both cases. When $\sigma_{\text e}=0.6$~m, about 90\% of the RP estimation errors are within $0.61$~m for IA-TS-RPR, while it is around $0.80$~m for the scheme proposed in~\cite{Xie2024ToA}. This corresponds to a reduction of $0.19$~m and an improvement of $23.8\%$. When $\sigma_{\text e}=1.2$~m, it becomes $0.73$~m and $1.40$~m for IA-TS-RPR and the scheme proposed in~\cite{Xie2024ToA}, respectively, which corresponds to a reduction of $0.67$~m and an improvement of $47.9\%$. These results show that IA-TS-RPR maintains sub-meter RP localization accuracy under severe user positioning error, which further confirms the superior robustness of IA-TS-RPR.

The scheme proposed in \cite{Xie2024ToA} relies on ToA-based ranging, which requires stringent time synchronization and is highly sensitive to timing biases. As a result, its estimation performance degrades significantly when the user localization error increases. In contrast, by leveraging the physical characteristic that a single reflector corresponds to a unique VBS, and integrating sensing-area-based information aggregation, the proposed IA-TS-RPR algorithm effectively fuses estimation results from multiple propagation paths to extract environmental structural features. This enhances the noise suppression capability and leads to significant improvements in estimation accuracy. }

\subsection{Performance Analysis of Environmental Variation Sensing}
To evaluate the effectiveness of the proposed environmental change sensing, this work analyzes the distribution of Hausdorff distances between the real-time VBS sets perceived by users and the corresponding VBS sets stored in the original CKM (associated with Map 0) under four environmental variation scenarios (Map 1–4). We set $B=10$ and ${{\varepsilon }_{D}}=0.6$ in the proposed EVS-HC-TWV algorithm. 


As shown in Fig.~\ref{fig3.11}, most Hausdorff distance values are concentrated near zero, indicating that the VBSs at user positions are largely consistent with those at the corresponding CKM GPs, i.e., no significant environmental changes have occurred. However, certain user positions exhibit relatively large Hausdorff distances, and these positions are critical for detecting environmental variation and determining RP positions. Therefore, the magnitude of the Hausdorff distance provides a simple and efficient means for identifying whether environmental changes have taken place.
\begin{figure}[!thb]
	\centering
	{\includegraphics[width=\linewidth]{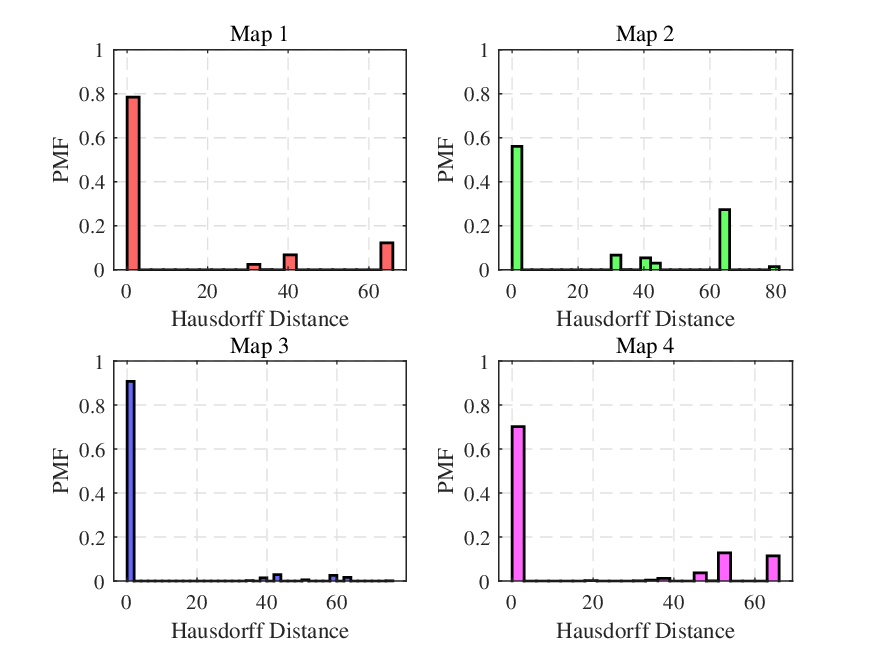}}
	\caption{Probability mass function (PMF) of the VBS Hausdorff distance for Maps 1-4 with respect to Map 0.}
	\label{fig3.11}
\end{figure}

Fig.~\ref{fig3.12} illustrates the anomalous GUs identified (left) and the estimated positions of changed RPs (right)  using the EVS-HC-TWV algorithm for the four scenarios. As shown in the left panels, the entire simulation area is uniformly divided into GUs with 1$\,$m spacing. Brown regions indicate static buildings, green regions represent anomalous GUs, and red regions denote the environment variation areas identified by the proposed EVS-HC-TWV algorithm. It can be observed from Fig.~\ref{fig3.12} that each scenario contains several newly introduced reflectors with distinct position distributions. In all cases, the proposed method successfully senses the locations of the newly added reflectors.
A specific example of Map 2 is shown in Fig.~\ref{fig3.13}. It can be seen that  due to the presence of reflector D, signal propagation paths in region 1 are blocked, while regions 2 and 3 experience newly introduced reflected paths. The proposed EVS-HC-TWV is capable of accurately identifying these anomalies by sensing the variation in propagation paths and thereby precisely localizing the position of the changed reflector.

\begin{figure}[!thb]
	\centering
	{\includegraphics[width=.9\linewidth]{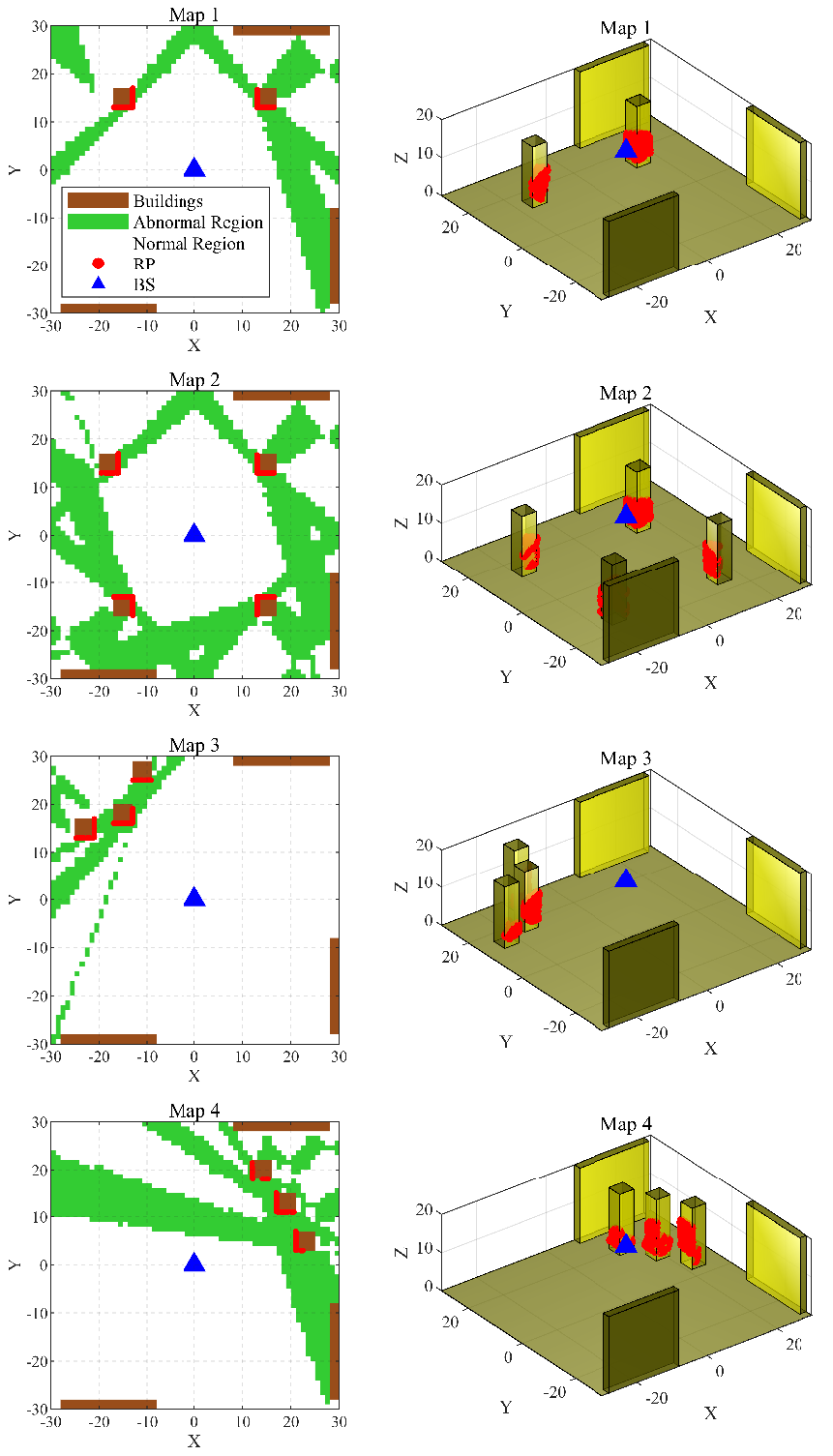}}
	\caption{Illustration of anomalous GUs (left)  and estimated positions of the changed reflectors (right).}
	\label{fig3.12}
\end{figure}

\begin{figure}[!thb]
	\centering
	{\includegraphics[width=0.7\linewidth]{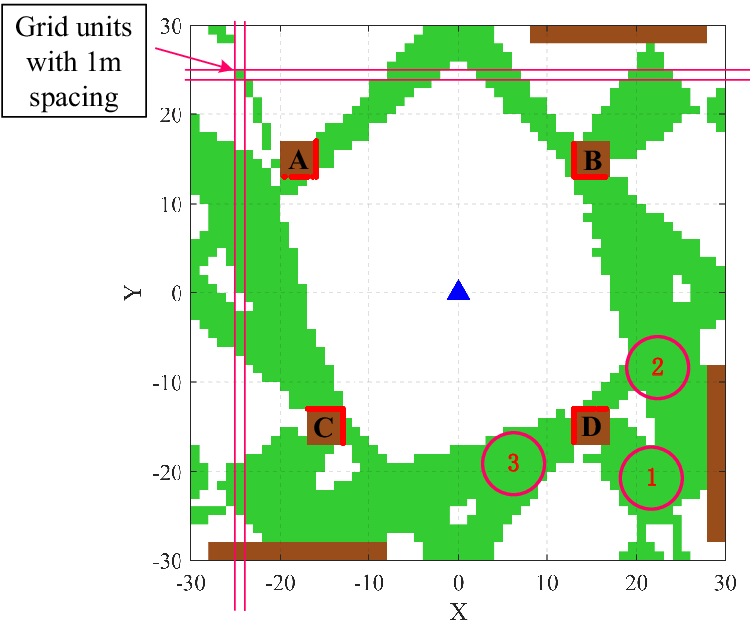}}
	\caption{A specific example of  Map 2,  where the map is divided into grids with a spacing of 1 m. The labels A, B, C, and D indicate newly added reflectors, and 1, 2, 3 denote the detected abnormal region indices.}
	\label{fig3.13}
\end{figure}

The detailed analysis of Map 2 in Fig. \ref{fig3.13}  clearly demonstrates the high accuracy and robustness of the proposed algorithm in complex environments. Specifically, the algorithm can precisely capture the impact of changed reflectors on channel propagation characteristics and promptly make accurate judgments regarding anomalous environmental changes. Similar phenomena and conclusions are also validated by the simulation results of Map 1, Map 3, and Map 4. Despite the differences in the specific anomalous regions and RP configurations, the proposed algorithm consistently maintains high detection accuracy and stability, further confirming its generalizability and robustness.

{\color{black}

{\color{black}Table~\ref{tab:acc_gp_user_perfect_delta3} reports the sensing accuracy under different amounts of collected user observations $K$ and different CKM grid spacings $d_{\mathrm{ckm}}$, and compares the perfect case with a noisy case. Here, $K$ denotes the number of user observations collected within the considered sensing window for the proposed EVS-HC-TWV scheme.} Following (\ref{equ1.9}), we construct the noisy case by adding random noise with $\sigma^2=3$ to the angle observations. The results show that the accuracy is close to saturation when the CKM grid is dense. For example, with $K=500$ and $d_{\mathrm{ckm}}=1\,\mathrm{m}$, the accuracy reaches $99.97\%$ in the perfect case and remains $99.94\%$ in the noisy case, indicating that a higher-resolution prior improves the HNN matching precision and stabilizes the subsequent decision. As $d_{\mathrm{ckm}}$ increases, the spatial quantization error becomes larger and the performance degrades slightly. For a fixed $d_{\mathrm{ckm}}$, the noisy case introduces an additional degradation compared with the perfect case, and the degradation is more evident with a coarser grid and smaller user redundancy. In general, the SA-level aggregation and the TWV mechanism leverage multi-user and multi-slot redundancy to suppress outliers and occasional deviations. Therefore, the overall accuracy remains high and stable even with a relatively sparse grid in the noisy case, demonstrating robustness to angular errors and CKM grid settings.
\begin{table}[t]
{\color{black}
\centering
\footnotesize 
\caption{Accuracy of environmental variation region sensing (\%) versus the number of users $K$ under different CKM grid spacings $d_\mathrm{ckm}$ (perfect/noisy).}
\label{tab:acc_gp_user_perfect_delta3}

\renewcommand{\arraystretch}{1.12}
\setlength{\tabcolsep}{5.0pt}

\begin{adjustbox}{max width=\textwidth,center}
\begin{tabular}{c ccc}
\toprule
\diagbox{$K$}{$d_{\mathrm{ckm}}$}
	& 1 m  & 3 m  & 5 m \\
	\midrule
	500    & 99.97 / 99.94  & 99.00 / 98.75  & 97.92 / 95.83 \\
	1000   & 99.97 / 99.92  & 98.75 / 98.00  & 98.61 / 96.53 \\
	5000   & 99.89 / 99.81  & 99.50 / 99.25  & 98.61 / 97.92 \\
	10000  & 99.83 / 99.67  & 99.75 / 99.25  & 97.92 / 97.22 \\
	\bottomrule
\end{tabular}
\end{adjustbox}}
\end{table}

{\color{black} The robustness of the proposed EVS-HC-TWV scheme against practical multipath observation imperfections including false alarm and missed detections is evaluated in Fig.~\ref{CKM2_path_response}, where the imperfection ratio denotes the proportion of affected paths among all observed paths. This ratio is set to 3\%, 5\%, 10\%, and 15\%, respectively. For each setting, 50 independent trials are conducted with $K=500$ collected user observations under three CKM grid spacings, i.e., $d_{\mathrm{ckm}}=1$ m, 3 m, and 5 m. It can be observed that the sensing accuracy generally decreases as the imperfection ratio increases, and the degradation becomes more evident with a larger CKM grid spacing. When $d_{\mathrm{ckm}}=1$ m, the accuracy remains close to 99\% even under a 15\% imperfection ratio. When $d_{\mathrm{ckm}}=5$ m, the median accuracy decreases from above 90\% under low imperfection ratios to about 83\% at the 15\% imperfection ratio. Overall, the proposed scheme maintains high sensing accuracy under different imperfection ratios, verifying the robustness and accuracy of EVS-HC-TWV against practical multipath observation imperfections.
\begin{figure}[!thb]
    \centering
    \includegraphics[width=.8\linewidth]{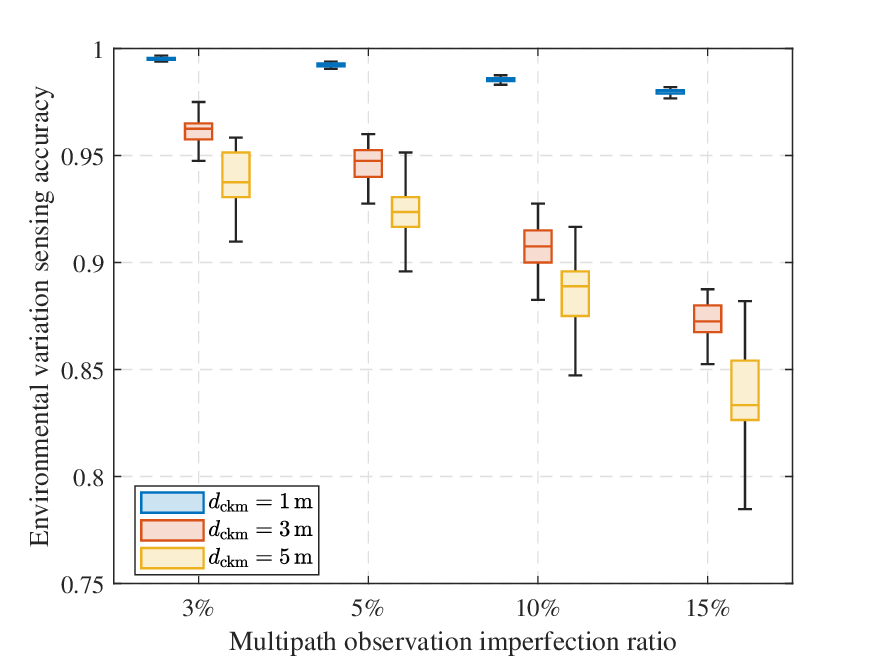}
    \caption{Box plots of environmental variation sensing accuracy under different multipath observation imperfection ratios.}
    \label{CKM2_path_response}
\end{figure}
}

{\color{black}To further validate the effectiveness of the proposed EVS-HC-TWV in a real scene, we modeled a $150\times250~\mathrm{m}^2$ area at the campus of Jiangsu University, where the BS is deployed at $[0,0,15]$ and both appearing and disappearing reflectors are considered. To further examine the applicability to reflectors with different geometries, the validation scenario includes a polygonal appearing reflector and a circular disappearing reflector. There are $K=1000$ users and the CKM grid spacing $d_{\mathrm{ckm}}=10~\mathrm{m}$. As shown in Fig.~\ref{realmap_com}, the sensing results of the proposed EVS-HC-TWV are spatially consistent with the baseline markers and show highly accurate RP localization. Moreover, the proposed scheme can further determine whether the RP is appearing or disappearing. The proposed EVS-HC-TWV attains an overall sensing accuracy of $96.27\%$ in this case, demonstrating its strong robustness and stable performance in realistic propagation environments.}

}

\begin{figure}[!thb]
	\centering
	{\includegraphics[width=\linewidth]{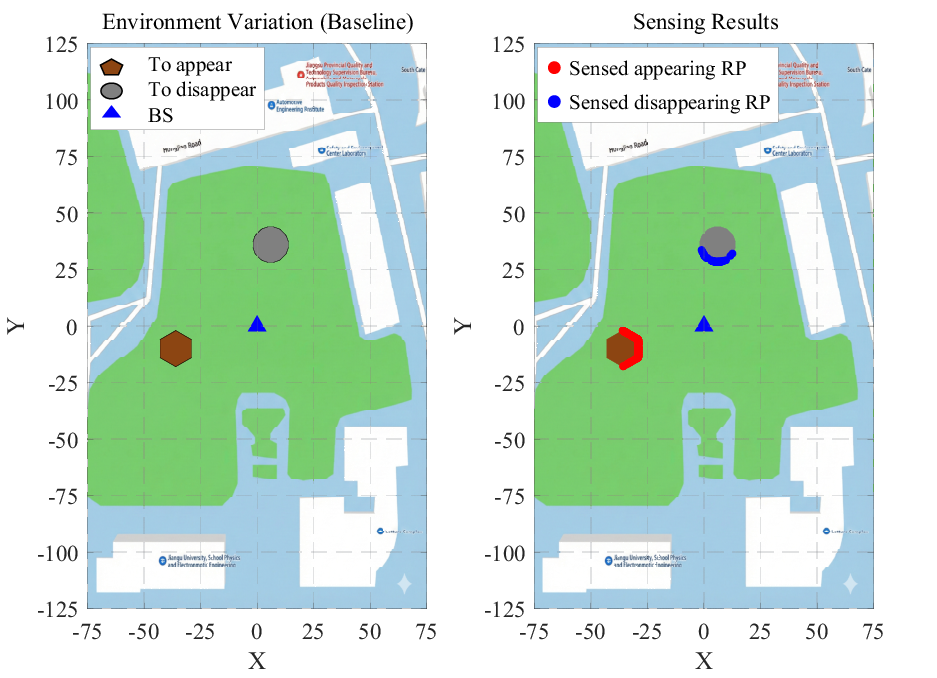}}
    {\color{black}
	\caption{Part of the campus at Jiangsu University for algorithm validation. Left): the baseline environment with one appearing and one disappearing target, respectively. Right): Sensing results of the proposed EVS-HC-TWV.}
	\label{realmap_com}}
\end{figure}

\subsection{Complexity Analysis}
The proposed CKM-enabled environment variation sensing algorithm is organized into four stages, and the computational complexity of each stage is summarized together with the measured CPU time in Table~\ref{table_complexity_time}. The proposed algorithm is tested on a platform with an Intel Core i5-13600KF CPU, where the number of users $K=250$ and CKM grid spacing $d_{\mathrm{ckm}}=1\,\mathrm{m}$. Here, $C_{\mathrm{match}}$, $N_{\mathrm{GU}}$ and $M_{\mathrm{abn}}$ denote the per-candidate matching cost, the number of GUs and that of the detected anomalous RPs, respectively. Denoting $T_{\mathrm{off}}$ and $T_{\mathrm{on}}$ as the offline and online CPU time, respectively, it can be concluded that $T_{\mathrm{off}}\propto N_{\mathrm{CKM}},\,T_{\mathrm{on}}\propto K$ and $N_{\mathrm{CKM}}\propto d_{\mathrm{ckm}}^{-2}$. The results show that the overall online overhead remains at the sub-second level in this case. 

\begin{table}[!thb]
		\centering
		\caption{Computational complexity of the proposed scheme.}
		\label{table_complexity_time}
        \small
		\begin{tabular}{m{2cm}lm{1.3cm}}
			\toprule[1.5pt]
			Stage & Complexity & CPU time \\
			\midrule[0.75pt]
			Offline: CKM preprocessing
			& $\mathcal{O}\!\left(\left(T_{\text{max}}+N_{\text{max}}\right)\sum_{n=1}^{N_{\text{CKM}}} L_n\right)$
			& 2.27s \\
			
			Online: user information processing
			& $\mathcal{O}\!\left((T_{\text{max}}+N_{\text{max}})\sum_{k=1}^{K}L_k\right)$
			& 0.17s \\
			
			Online: user-GP matching
			& $\mathcal{O}\!\left(K\left(N_\text{H}\log N_{\text{CKM}}+N_\text{H} C_{\text{match}}\right)\right)$
			& 0.15s \\
			
			Online: voting \& anomalous RPs extraction
			& $\mathcal{O}\!\left(K+N_{\text{GU}}+M_{\text{abn}}\right)$
			& 0.004s \\
			\bottomrule[1.5pt]
		\end{tabular}
\end{table}

{\color{black}
\subsection{Limitations and Future Work}
 First, the current framework focuses on single-reflection NLoS paths, and more complex propagation mechanisms, such as multi-bounce reflections, can be incorporated to improve its applicability in richer propagation environments. Second, the proposed method relies on the consistency between the historical CKM and the current propagation environment. Although the sensitivity to CKM grid spacing has been evaluated, CKM maintenance and online updating can be further investigated to reduce the influence of outdated or spatially misaligned CKMs on user-GP association and VBS-set comparison. Third, this work mainly considers reflector-position variations, including appearing and disappearing reflectors. The joint sensing of reflector positions and other physical attributes, such as size, material type, and reflection characteristics, can also be explored in future work.}

\section{Conclusion}
This work has managed to address the problem of dynamic environmental sensing caused by environment variation and has proposed a CKM-based environmental variation sensing framework. First, a probabilistic model is formulated to characterize the CKM-driven environmental variation sensing problem. Then, a two-stage RP estimation algorithm, IA-TS-RPR, is developed based on SA information aggregation. Finally, by leveraging VBS distribution feature sets, an EVS-HC-TWV algorithm is proposed to enhance robustness in dynamic environments. Simulation results validate the effectiveness and robustness of the proposed approach. The IA-TS-RPR algorithm achieves sub-meter level RP position estimation accuracy under noisy conditions, offering substantial improvements over the ToA-based scheme \cite{Xie2024ToA}, LMMSE and LS. Furthermore, the EVS-HC-TWV algorithm accurately localizes environment variation regions while maintaining overall sensing accuracy above 95\%, reaching 99\% in specific environments.

\small
\bibliographystyle{IEEEtran}
\bibliography{reference}

@ARTICLE{Wu2026TWC,
  author={Wu, Di and Dai, Zhuoyin and Zeng, Yong},
  journal={IEEE Transactions on Wireless Communications}, 
  title={You May Use the Same Channel Knowledge Map for Environment-Aware {NLoS} Sensing and Communication}, 
  year={2026},
  volume={25},
  number={},
  pages={14627-14641},
  doi={10.1109/TWC.2026.3678277}}

@ARTICLE{ref2_ISAC,
	author={Liu, An and Huang, Zhe and Li, Min and Wan, Yubo and Li, Wenrui and Han, Tony Xiao and Liu, Chenchen and Du, Rui and Tan, Danny Kai Pin and Lu, Jianmin and Shen, Yuan and Colone, Fabiola and Chetty, Kevin},
	journal={IEEE Communications Surveys \& Tutorials}, 
	title={A Survey on Fundamental Limits of Integrated Sensing and Communication}, 
	year={2022},
	volume={24},
	number={2},
	pages={994-1034},
	doi={10.1109/COMST.2022.3149272}}

@ARTICLE{ref3_ISAC,
	author={Liu, Rang and Li, Ming and Luo, Honghao and Liu, Qian and Swindlehurst, A. Lee},
	journal={IEEE Wireless Communications}, 
	title={Integrated Sensing and Communication with Reconfigurable Intelligent Surfaces: Opportunities, Applications, and Future Directions}, 
	year={2023},
	volume={30},
	number={1},
	pages={50-57},
	doi={10.1109/MWC.002.2200206}}

@ARTICLE{ref4_ISAC,
	author={Liu, Fan and Cui, Yuanhao and Masouros, Christos and Xu, Jie and Han, Tony Xiao and Eldar, Yonina C. and Buzzi, Stefano},
	journal={IEEE Journal on Selected Areas in Communications}, 
	title={Integrated Sensing and Communications: Toward Dual-Functional Wireless Networks for {6G} and Beyond}, 
	year={2022},
	volume={40},
	number={6},
	pages={1728-1767},
	doi={10.1109/JSAC.2022.3156632}}

@ARTICLE{ref5_ISAC,
	author={Lotti, Marina and Pasolini, Gianni and Guerra, Anna and Guidi, Francesco and D'Errico, Raffaele and Dardari, Davide},
	journal={IEEE Journal of Selected Topics in Signal Processing}, 
	title={Radio {SLAM} for {6G} Systems at {THz} Frequencies: Design and Experimental Validation}, 
	year={2023},
	volume={17},
	number={4},
	pages={834-849},
	doi={10.1109/JSTSP.2023.3285101}}

@ARTICLE{ref6_ISAC,
	author={Chen, Jie and Wang, Xianbin},
	journal={IEEE Journal of Selected Topics in Signal Processing}, 
	title={Learning-Based Intermittent {CSI} Estimation With Adaptive Intervals in Integrated Sensing and Communication Systems}, 
	year={2024},
	volume={18},
	number={5},
	pages={917-932},
	doi={10.1109/JSTSP.2024.3468037}}

@ARTICLE{ref8_ISAC,
	author={Tang, Lun and Wang, Asha and Xia, Bingsen and Tang, Yuanchun and Chen, Qianbin},
	journal={IEEE Internet of Things Journal}, 
	title={Research on Integrated Sensing, Communication Resource Allocation, and Digital Twin Placement Based on Digital Twin in {IoV}}, 
	year={2025},
	volume={12},
	number={11},
	pages={17300-17315},
	doi={10.1109/JIOT.2025.3535767}}

@article{ref10_ISAC,
  title={A tutorial on {MIMO-OFDM ISAC}: From far-field to near-field},
  author={Dai, Qianglong and Zeng, Yong and Wang, Huizhi and You, Changsheng and Zhou, Chao and Cheng, Hongqiang and Xu, Xiaoli and Jin, Shi and Swindlehurst, A Lee and Eldar, Yonina C and others},
  journal={IEEE Communications Surveys \& Tutorials},
  volume={28},
  pages={4319--4358},
  year={2026},
  publisher={IEEE}
}

@ARTICLE{ref2_CKM,
  author={Xu, Zihan and Zhou, Zhiwen and Wu, Di and Xu, Xiaoli and Zeng, Yong},
  journal={IEEE Internet of Things Journal}, 
  title={{CKM}-Enabled Joint Spatial--Doppler Domain Clutter Suppression for Low-Altitude {UAV} {ISAC}}, 
  year={2026},
  volume={13},
  number={14},
  pages={31752-31767},
  doi={10.1109/JIOT.2026.3689485}}

@ARTICLE{ref3_CKM,
	author={Zhao, Tianxiao and Wang, Xin and Yang, Wenfei and Xi, Xiaojun and Li, Jian},
	journal={IEEE Communications Letters}, 
	title={On the Common {AOA} Error in {CKM}-Based Integrated Sensing and Communications}, 
	year={2023},
	volume={27},
	number={7},
	pages={1859-1863},
	doi={10.1109/LCOMM.2023.3271356}}

@ARTICLE{ref4_CKM,
	author={Zeng, Yong and Xu, Xiaoli},
	journal={IEEE Wireless Communications}, 
	title={Toward Environment-Aware {6G} Communications via Channel Knowledge Map}, 
	year={2021},
	volume={28},
	number={3},
	pages={84-91},
	doi={10.1109/MWC.001.2000327}}

@ARTICLE{ref5_CKM,
	author={Zeng, Yong and Chen, Junting and Xu, Jie and Wu, Di and Xu, Xiaoli and Jin, Shi and Gao, Xiqi and Gesbert, David and Cui, Shuguang and Zhang, Rui},
	journal={IEEE Communications Surveys \& Tutorials}, 
	title={A Tutorial on Environment-Aware Communications via Channel Knowledge Map for {6G}}, 
	year={2024},
	volume={26},
	number={3},
	pages={1478-1519},
	doi={10.1109/COMST.2024.3364508}}

@ARTICLE{rel2_SLAM,
	author={Gentner, Christian and Jost, Thomas and Wang, Wei and Zhang, Siwei and Dammann, Armin and Fiebig, Uwe-Carsten},
	journal={IEEE Transactions on Wireless Communications}, 
	title={Multipath Assisted Positioning with Simultaneous Localization and Mapping}, 
	year={2016},
	volume={15},
	number={9},
	pages={6104-6117},
	doi={10.1109/TWC.2016.2578336}}

@ARTICLE{rel3_SLAM,
	author={Leitinger, Erik and Venus, Alexander and Teague, Bryan and Meyer, Florian},
	journal={IEEE Transactions on Signal Processing}, 
	title={Data Fusion for Multipath-Based {SLAM}: Combining Information From Multiple Propagation Paths}, 
	year={2023},
	volume={71},
	number={},
	pages={4011-4028},
	doi={10.1109/TSP.2023.3310360}}

@ARTICLE{rel4_SLAM,
	author={Leitinger, Erik and Meyer, Florian and Hlawatsch, Franz and Witrisal, Klaus and Tufvesson, Fredrik and Win, Moe Z.},
	journal={IEEE Transactions on Wireless Communications}, 
	title={A Belief Propagation Algorithm for Multipath-Based {SLAM}}, 
	year={2019},
	volume={18},
	number={12},
	pages={5613-5629},
	doi={10.1109/TWC.2019.2937781}}

@INPROCEEDINGS{rel5_SLAM,
	author={Wielandner, Lukas and Venus, Alexander and Wilding, Thomas and Leitinger, Erik},
	booktitle={2023 26th International Conference on Information Fusion (FUSION)}, 
	title={Multipath-based {SLAM} with Multiple-Measurement Data Association}, 
	year={2023},
	volume={},
	number={},
	pages={1-8},
	doi={10.23919/FUSION52260.2023.10224195}}

@ARTICLE{rel8_SLAM,
	author={Yang, Jie and Wen, Chao-Kai and Jin, Shi and Li, Xiao},
	journal={IEEE Transactions on Wireless Communications}, 
	title={Enabling Plug-and-Play and Crowdsourcing {SLAM} in Wireless Communication Systems}, 
	year={2022},
	volume={21},
	number={3},
	pages={1453-1468},
	doi={10.1109/TWC.2021.3104006}}

@INPROCEEDINGS{rel3_CKM,
	author={Mo, Ronghong and Pei, Yiyang and Sun, Sumei and Premkumar, A. B. and Venkatarayalu, Neelakantam V},
	booktitle={2024 IEEE 99th Vehicular Technology Conference (VTC2024-Spring)}, 
	title={Deep Machine Learning-Based {AoD} Map and {AoA} Map Construction for Wireless Networks}, 
	year={2024},
	volume={},
	number={},
	pages={1-5},
	doi={10.1109/VTC2024-Spring62846.2024.10683141}}

@ARTICLE{rel1_CKMY,
	author={Wu, Di and Zeng, Yong and Jin, Shi and Zhang, Rui},
	journal={IEEE Transactions on Wireless Communications}, 
	title={Environment-Aware Hybrid Beamforming by Leveraging Channel Knowledge Map}, 
	year={2024},
	volume={23},
	number={5},
	pages={4990-5005},
	doi={10.1109/TWC.2023.3323941}}

@ARTICLE{rel6_CKMY,
	author={Li, Haoyun and Li, Peiming and Cheng, Gaoyuan and Xu, Jie and Chen, Junting and Zeng, Yong},
	journal={Journal of Communications and Information Networks}, 
	title={Channel Knowledge Map ({CKM})-Assisted Multi-{UAV} Wireless Network: {CKM} Construction and {UAV} Placement}, 
	year={2023},
	volume={8},
	number={3},
	pages={256-270},
	doi={10.23919/JCIN.2023.10272353}}

@ARTICLE{rel7_CKMY,
  author={Zhang, Chaoyue and Zhou, Zhiwen and Xu, Xiaoli and Zeng, Yong and Zhang, Zaichen and Jin, Shi},
  journal={IEEE Transactions on Vehicular Technology}, 
  title={Prototyping and Experimental Results for {ISAC}-Based Channel Knowledge Map}, 
  year={2025},
  volume={74},
  number={7},
  pages={10719-10731},
  doi={10.1109/TVT.2025.3545785}}

@article{rel9_CKMY,
  author = {Fu, Shen and Zeng, Yong and Wu, Zijian and Wu, Di and Xu, Jie and Jin, Shi and Wang, Cheng-Xiang and Gao, Xiqi},
  title = {{CKMDiff}: a generative diffusion model for {CKM} construction via inverse problems with learned priors},
  journal = {npj Wireless Technology},
  year = {2026},
  volume = {2},
  number = {1},
  pages = {35},
  doi = {10.1038/s44459-026-00042-1},
}

@ARTICLE{ref1_BCD,
	author={Li, Jia and Xiao, Mingqing and Fang, Cong and Dai, Yue and Xu, Chao and Lin, Zhouchen},
	journal={IEEE Transactions on Pattern Analysis and Machine Intelligence}, 
	title={Training Neural Networks by Lifted Proximal Operator Machines}, 
	year={2022},
	volume={44},
	number={6},
	pages={3334-3348},
	doi={10.1109/TPAMI.2020.3048430}}

@ARTICLE{ref2_BCD,
	author={Pan, Cunhua and Ren, Hong and Wang, Kezhi and Xu, Wei and Elkashlan, Maged and Nallanathan, Arumugam and Hanzo, Lajos},
	journal={IEEE Transactions on Wireless Communications}, 
	title={Multicell {MIMO} Communications Relying on Intelligent Reflecting Surfaces}, 
	year={2020},
	volume={19},
	number={8},
	pages={5218-5233},
	doi={10.1109/TWC.2020.2990766}}

@ARTICLE{ref3_Armijo,
	author={Liu, Liqing and Chang, Zheng and Guo, Xijuan},
	journal={IEEE Internet of Things Journal}, 
	title={Socially Aware Dynamic Computation Offloading Scheme for Fog Computing System With Energy Harvesting Devices}, 
	year={2018},
	volume={5},
	number={3},
	pages={1869-1879},
	doi={10.1109/JIOT.2018.2816682}}

@article{ref_gqd,
  title={Channel reconstruction for {mmWave} Massive {MIMO} systems based on Channel Path Map},
  author={Gao, Qidong and Li, Zeyang and Zhang, Wence and Bao, Xu and Xia, Jing and Zheng, Zhaowen},
  journal={Physical Communication},
  volume={61},
  pages={102232},
  year={2023},
  publisher={Elsevier}
}

@INPROCEEDINGS{ref_V,
  author={Xinning Wei and Palleit, Nico and Weber, Tobias},
  booktitle={2011 IEEE 22nd International Symposium on Personal, Indoor and Mobile Radio Communications}, 
  title={{AOD/AOA/TOA}-based {3D} positioning in {NLOS} multipath environments}, 
  year={2011},
  volume={},
  number={},
  pages={1289-1293},
  doi={10.1109/PIMRC.2011.6139709}}

@ARTICLE{ref_6g,
  author={Saad, Walid and Bennis, Mehdi and Chen, Mingzhe},
  journal={IEEE Network}, 
  title={A Vision of {6G} Wireless Systems: Applications, Trends, Technologies, and Open Research Problems}, 
  year={2020},
  volume={34},
  number={3},
  pages={134-142},
  doi={10.1109/MNET.001.1900287}}

@ARTICLE{ref_SV,
  author={Meijerink, Arjan and Molisch, Andreas F.},
  journal={IEEE Transactions on Antennas and Propagation}, 
  title={On the Physical Interpretation of the Saleh–Valenzuela Model and the Definition of Its Power Delay Profiles}, 
  year={2014},
  volume={62},
  number={9},
  pages={4780-4793},
  doi={10.1109/TAP.2014.2335812}}

@article{ref_why,
title = {Localization performance analysis based on channel knowledge map},
journal = {Physical Communication},
volume = {72},
pages = {102721},
year = {2025},
doi = {https://doi.org/10.1016/j.phycom.2025.102721},
author = {Huayang Wei and Qidong Gao and Wence Zhang and Jing Xia and Zhaowen Zheng and Xu Bao}
}

@ARTICLE{ref_LS,
  author={Balevi, Eren and Andrews, Jeffrey G.},
  journal={IEEE Transactions on Wireless Communications}, 
  title={Wideband Channel Estimation With a Generative Adversarial Network}, 
  year={2021},
  volume={20},
  number={5},
  pages={3049-3060},
  doi={10.1109/TWC.2020.3047100}}

@ARTICLE{user_error1,
  author={Que, Hang and Yang, Jie and Du, Tao and Xia, Shuqiang and Wen, Chao-Kai and Jin, Shi},
  journal={IEEE Transactions on Communications}, 
  title={Cooperative Mapping, Localization, and Beam Management via Multi-Modal {SLAM} in {ISAC} Systems}, 
  year={2025},
  volume={73},
  number={11},
  pages={12023-12037},
  doi={10.1109/TCOMM.2025.3588548}}

@ARTICLE{user_error2,
  author={Chen, Yilong and Ren, Zixiang and Xu, Jie and Zeng, Yong and Ng, Derrick Wing Kwan and Cui, Shuguang},
  journal={IEEE Communications Magazine}, 
  title={Integrated Sensing, Communication, and Powering: Toward Multi-Functional {6G} Wireless Networks}, 
  year={2025},
  volume={63},
  number={8},
  pages={146-153},
  doi={10.1109/MCOM.006.2400013}}

@ARTICLE{Xie2024ToA,
  author={Xie, Jincheng and Luo, Baojia and Dong, Miaomiao and Chen, Xiang and Han, Wei},
  journal={IEEE Communications Letters}, 
  title={Environment Topology Oriented Fingerprint Localization in Indoor {NLoS} Scenarios}, 
  year={2024},
  volume={28},
  number={3},
  pages={523-527},
  doi={10.1109/LCOMM.2024.3357740}}

@ARTICLE{venus2024graph,
  author={Venus, Alexander and Leitinger, Erik and Tertinek, Stefan and Witrisal, Klaus},
  journal={IEEE Transactions on Wireless Communications}, 
  title={A Graph-Based Algorithm for Robust Sequential Localization Exploiting Multipath for Obstructed-{LOS}-Bias Mitigation}, 
  year={2024},
  volume={23},
  number={2},
  pages={1068-1084},
  doi={10.1109/TWC.2023.3285530}}

@ARTICLE{yan2025TGCN,
  author={Yan, Han and Chen, Hua and Liu, Wei and Yang, Songjie and Wang, Gang and Yuen, Chau},
  journal={IEEE Transactions on Green Communications and Networking}, 
  title={{RIS}-Enabled Joint Near-Field {3D} Localization and Synchronization in {SISO} Multipath Environments}, 
  year={2025},
  volume={9},
  number={1},
  pages={367-379},
  doi={10.1109/TGCN.2024.3422992}}

@ARTICLE{xiao2024TSP,
  author={Xiao, Zichao and Liu, Rang and Li, Ming and Liu, Qian and Swindlehurst, A. Lee},
  journal={IEEE Transactions on Signal Processing}, 
  title={A Novel Joint Angle-Range-Velocity Estimation Method for {MIMO-OFDM ISAC} Systems}, 
  year={2024},
  volume={72},
  number={},
  pages={3805-3818},
  doi={10.1109/TSP.2024.3442886}}

@ARTICLE{Keskin2024TWC,
  author={Keskin, Musa Furkan and Marcus, Carina and Eriksson, Olof and Alvarado, Alex and Widmer, Joerg and Wymeersch, Henk},
  journal={IEEE Transactions on Wireless Communications}, 
  title={Integrated Sensing and Communications With {MIMO-OTFS: ISI/ICI} Exploitation and Delay-Doppler Multiplexing}, 
  year={2024},
  volume={23},
  number={8},
  pages={10229-10246},
  doi={10.1109/TWC.2024.3370501}}

@ARTICLE{hu2024TWC,
  author={Hu, Yanmo and Wu, Kai and Zhang, J. Andrew and Deng, Weibo and Guo, Y. Jay},
  journal={IEEE Transactions on Wireless Communications}, 
  title={Performance Bounds and Optimization for {CSI}-Ratio-Based Bi-Static Doppler Sensing in {ISAC} Systems}, 
  year={2024},
  volume={23},
  number={11},
  pages={17461-17477},
  doi={10.1109/TWC.2024.3453902}}

@ARTICLE{Pegoraro2024TWC,
  author={Pegoraro, Jacopo and Lacruz, Jesus O. and Azzino, Tommy and Mezzavilla, Marco and Rossi, Michele and Widmer, Joerg and Rangan, Sundeep},
  journal={IEEE Transactions on Wireless Communications}, 
  title={JUMP: Joint Communication and Sensing With Unsynchronized Transceivers Made Practical}, 
  year={2024},
  volume={23},
  number={8},
  pages={9759-9775},
  doi={10.1109/TWC.2024.3365853}}

@ARTICLE{Li2025TVT,
  author={Li, Muye and Zhang, Shun and Ge, Yao and Yuen, Chau},
  journal={IEEE Transactions on Vehicular Technology}, 
  title={{STAR-RIS} Aided Dynamic Scatterers Tracking for Integrated Sensing and Communications}, 
  year={2025},
  volume={74},
  number={5},
  pages={7760-7773},
  doi={10.1109/TVT.2025.3527753}}

@ARTICLE{Baquero2022TVT,
  author={Baquero Barneto, Carlos and Rastorgueva-Foi, Elizaveta and Keskin, Musa Furkan and Riihonen, Taneli and Turunen, Matias and Talvitie, Jukka and Wymeersch, Henk and Valkama, Mikko},
  journal={IEEE Transactions on Vehicular Technology}, 
  title={Millimeter-Wave Mobile Sensing and Environment Mapping: Models, Algorithms and Validation}, 
  year={2022},
  volume={71},
  number={4},
  pages={3900-3916},
  doi={10.1109/TVT.2022.3146003}}

@ARTICLE{Shao2026IOT,
  author={Shao, Xiaoyan and Zhang, Wence and Li, Haohan and Shao, Zhichao and Zhang, Zhiguang and Bao, Xu},
  journal={IEEE Internet of Things Journal}, 
  title={A Low-Complexity Channel Knowledge Map Construction Based on Environmental Partitioning and Interpolation Weight Learning}, 
  year={2026},
  volume={13},
  number={11},
  pages={24044-24056},
  doi={10.1109/JIOT.2026.3672412}}

\vfill

\end{document}